\documentclass[a4paper]{article}
\usepackage{amsfonts}
\usepackage{amsmath}
\usepackage{amssymb}
\usepackage{mathtools}
\usepackage{calc}
\usepackage{tikz-cd}
\usepackage{braket}
\usepackage{fancybox}

\newcommand{\untaggedfootnote}[1]{\let\thefootnote\relax\footnote{#1}}

\makeatletter
\def\@crosshairs{\vbox to0pt{}}
\makeatother
\input{tcilatex}
\begin{document}

\title{Expectation and Price in Incomplete Markets}
\author{Paul McCloud \\
Department of Mathematics, University College London}
\maketitle

\begin{abstract}
Risk-neutral pricing dictates that the discounted derivative price is a
martingale in a measure equivalent to the economic measure. The residual
ambiguity for incomplete markets is here resolved by minimising the entropy
of the price measure from the economic measure, subject to mark-to-market
constraints, following arguments based on the optimisation of portfolio
risk. The approach accounts for market and funding convexities and
incorporates available price information, interpolating between
methodologies based on expectation and replication.

\untaggedfootnote{Author email: p.mccloud@ucl.ac.uk}
\end{abstract}

\vspace{3.5cm}

The principal innovation of the financial derivatives industry is the
ability to transform an arbitrary economic observable into a tradable
financial instrument. Provided that this is measurable at or prior to
settlement, the contractual terms dictate that the measurement of the
observable becomes the terminal cash value of the corresponding derivative.
Failure to deliver on the contract, and any other circumstances that alter
the settlement amount, are absorbed in the definition of the reference
observable, allowing default and external factors to be incorporated. This
contractualisation of economic observables permits the future exchange of
currencies in amounts that are undetermined at present.

The conceptual framework considered here assumes a universe of economic
observables whose values are revealed progressively through time, which are
then utilised as the cash settlement amounts for derivative securities in
one or more idealised currencies. Liquidity constraints are not considered,
and settlement is permitted in arbitrarily large positive or negative
amounts. No other properties of the securities are investigated, and
derivatives that match at settlement in all future scenarios are assumed to
be fungible.

\begin{figure}[!t]
\centering
\includegraphics[width=0.95%
\linewidth]{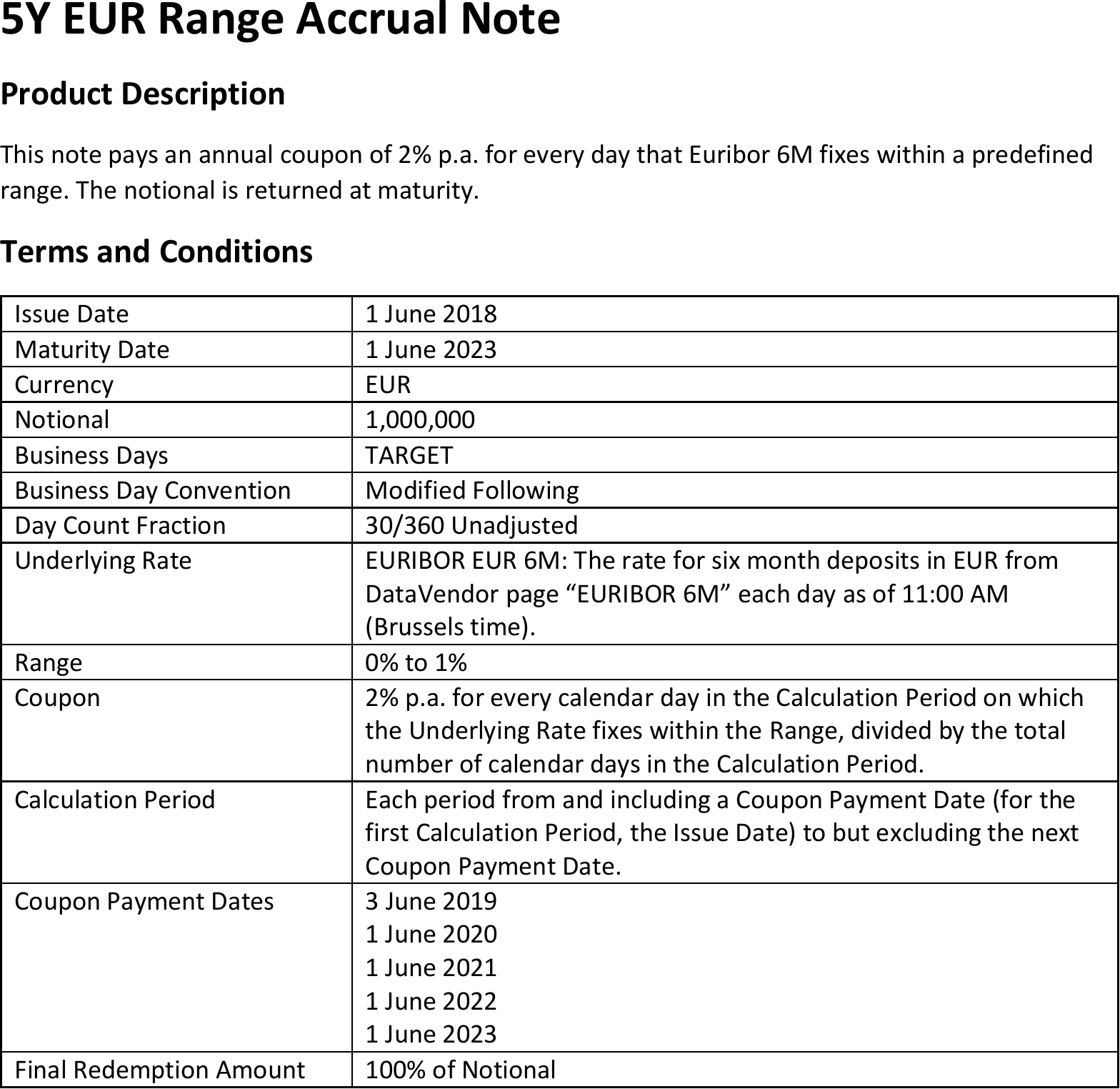}
\caption{A sample term sheet for a `range accrual' derivative. The term
sheet describes the observations (Euribor EUR 6M fixing) and calculations
(indicator function for a defined range) used to determine settlement
amounts, together with the dates on which the payments will be made.
Bilateral arrangements between the counterparties may prescribe further
payments, such as margin and collateral exchanges. The trade is also subject
to the terms of the legal jurisdiction, covering matters such as regulatory
capital requirements and actions in the event of default. All these
settlements related to the derivative potentially impact its valuation.}
\end{figure}

The price of the derivative must then account for \emph{discounting}, the
adjustment due to the funding of future settlements, and \emph{convexity},
the cost or benefit extracted from dynamic hedging in volatile markets.
Keeping these contributions in balance is the basis for the fair pricing of
the derivative. The profit/loss that the hedged derivative position accrues
over time is decomposed into two components:%
\begin{equation}
\text{Hedged P\&L}=\text{Carry}+\text{Gamma P\&L}
\end{equation}%
where carry is the change in the value of the hedged derivative arising from
funding costs and the decay in option value, and gamma p\&l is the residual
profit/loss due to unhedgeable convexity in the relationship between the
derivative price and the prices of liquid underlying securities used for
hedging. Efficient market dynamics reverts this net profit/loss toward zero,
leading to a balancing equation for the equilibrium price of the derivative.

\begin{figure}[!t]
\centering
\includegraphics[width=0.95%
\linewidth]{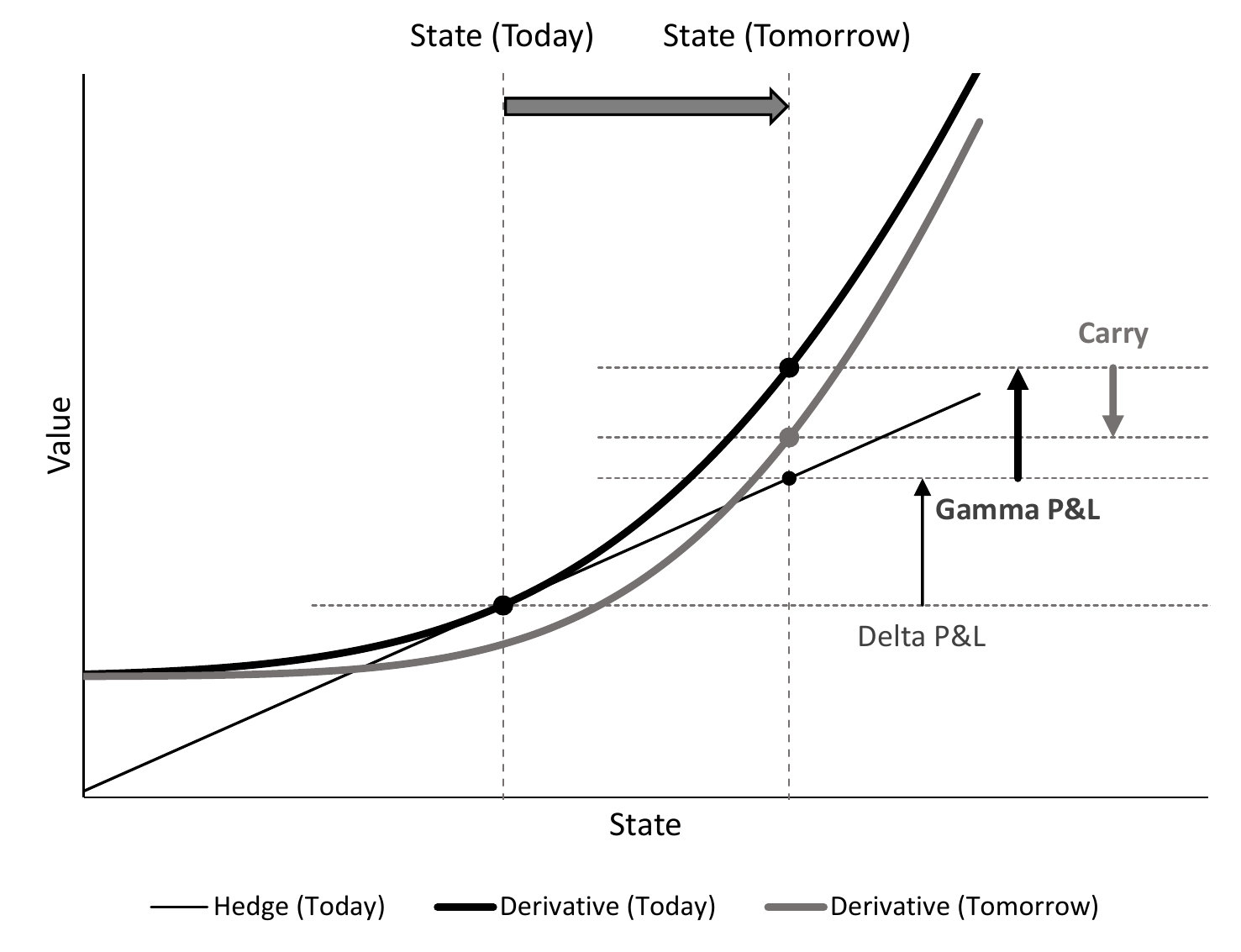}
\caption{Over a time interval, the profit/loss that accrues on the
derivative security is decomposed into the \emph{delta p\&l}, which is
hedged using liquid underlying securities, the \emph{gamma p\&l}, the
unhedgeable residual from the convexity of the price function, and \emph{%
carry}, the appreciation over time. Fair pricing requires that accrual from
gamma p\&l is offset by carry.}
\end{figure}

Terms representing carry and gamma p\&l are evident in the Black-Scholes
equation for the price $\mathsf{c}[t,\mathsf{s}]$ of a derivative security
contingent on the price $\mathsf{s}$ of an underlying security. In this
model, the underlying price is assumed to diffuse lognormally. Dynamic
hedging eliminates the market risk from the volatility of the underlying
price; setting the funded return to zero then leads to the equation:%
\begin{equation}
0=\frac{\partial \mathsf{c}}{\partial t}-r(\mathsf{c}-\mathsf{s}\frac{%
\partial \mathsf{c}}{\partial \mathsf{s}})+\frac{1}{2}\sigma ^{2}\mathsf{s}%
^{2}\frac{\partial ^{2}\mathsf{c}}{\partial \mathsf{s}^{2}}
\end{equation}%
for the derivative price, where $r$ is the funding rate and $\sigma $ is the
lognormal volatility of the underlying price. The derivative is delta-hedged
by an offsetting position in the underlying. Gamma p\&l accrues for positive
convexity when the underlying price is volatile, and the Black-Scholes
equation balances this `volatility$\times $convexity' accrual with the carry
from time decay and the funding costs of the hedged position.

Generalisations of the Black-Scholes model have equivalent terms for these
contributions, incorporating alternative volatility models for the
underlying prices and other economic variables. The viability of this
approach then depends on the effectiveness of the market risk transfer from
derivative to underlying provided by dynamic hedging, and the accuracy of
the model compared to realised market volatility over the lifetime of the
derivative.

\section{Economic principles for pricing}

The elementary economic principles underpinning this argument are
replicability and the absence of arbitrage, both essentially algebraic
constraints on the valuation map from the observable that represents the
settlement amount of the derivative to its price. As with all such
principles, they are an approximation to the reality of financial markets,
but their validity is assumed throughout the following.

The exposition rests on three founding economic principles:

\begin{description}
\item[The Principle of Replicability:] A security constructed as a linear
combination of underlying securities has price equal to the same linear
combination of underlying prices.

\item[The Principle of No-Arbitrage:] A security that has positive
settlement in all future scenarios has positive price.

\item[The Principle of Economic Equivalence:] A security that has zero
settlement in all possible future scenarios has zero price.
\end{description}

\noindent The first two principles are consistency conditions that disable
the construction of arbitrages, and dictate that the valuation map is linear
and positive. Prohibiting unattainable outcomes from impacting price, the
final principle connects the valuation map with the forecasting model of the
economy -- expectations of future outcomes are not necessarily reflected in
pricing, but this weaker requirement removes from consideration scenarios
that have zero measure in the economic model, and so are deemed to be
impossible. The connection between these economic principles and fundamental
mathematical constructions is the main focus of this essay.

\emph{The principles of replicability and no-arbitrage connect price with
the functional calculus of observables.} Let $\mathsf{A}$ be the space of
economic observables whose values can be ascertained at settlement.
Following the opening comments, the payoff of the derivative in a nominated
payment currency is represented as an observable in this space. The price
model is then an operation:%
\begin{equation}
\mathsf{z}:\mathsf{a}\in \mathsf{A}\mapsto \mathsf{z}\bullet \mathsf{a}\in 
\mathbb{R}
\end{equation}%
that maps the derivative payoff $\mathsf{a}$ to its price $\mathsf{z}\bullet 
\mathsf{a}$.

The principle of replicability imposes the homomorphic relationship with
addition:%
\begin{equation}
\mathsf{z}\bullet (\mathsf{a}+\mathsf{b})=(\mathsf{z}\bullet \mathsf{a})+(%
\mathsf{z}\bullet \mathsf{b})
\end{equation}%
thus ensuring that the price of a portfolio can be determined from the
prices of its constituents. The principle of no-arbitrage does not impose
the homomorphic relationship with multiplication, $\mathsf{z}\bullet (%
\mathsf{ab})=(\mathsf{z}\bullet \mathsf{a})(\mathsf{z}\bullet \mathsf{b})$,
as this denies the possibility of extracting value from convexity, an
essential property for a viable model of derivative prices. Instead, this
principle imposes the weaker requirement that the price map is positive:%
\begin{equation}
\mathsf{a}\geq 0\implies \mathsf{z}\bullet \mathsf{a}\geq 0
\end{equation}%
which in turn implies the sub-homomorphic relationship with multiplication:%
\begin{equation}
(\mathsf{z}\bullet \mathsf{ab})^{2}\leq (\mathsf{z}\bullet \mathsf{a}^{2})(%
\mathsf{z}\bullet \mathsf{b}^{2})
\end{equation}%
This Cauchy-Schwarz inequality for the price model is satisfied by linear
combinations of ring homomorphisms with positive weights.

\emph{The principle of economic equivalence connects price with the
stochastic calculus of observables.} The founding economic principles apply
across any time interval $t_{i}\leq t_{j}$ with price model given by a
linear and positive operation:%
\begin{equation}
\mathsf{z}_{ij}:\mathsf{a}_{j}\in \mathsf{A}_{j}\mapsto \mathsf{z}%
_{ij}\bullet \mathsf{a}_{j}\in \mathsf{A}_{i}
\end{equation}%
that maps the derivative payoff $\mathsf{a}_{j}$ at time $t_{j}$ to its
price $\mathsf{z}_{ij}\bullet \mathsf{a}_{j}$ at time $t_{i}$, where $%
\mathsf{A}_{i}\subset \mathsf{A}_{j}$ are the subspaces of observables whose
values can be ascertained at the start and end of the interval.

The principle of economic equivalence relates these operations to the model
of the economy:%
\begin{equation}
\mathsf{z}_{ij}\bullet \mathsf{a}_{j}=\mathsf{\bar{z}}_{ij}\bullet \mathsf{w}%
_{ij}\mathsf{a}_{j}
\end{equation}%
where:%
\begin{equation}
\mathsf{\bar{z}}_{ij}:\mathsf{a}_{j}\in \mathsf{A}_{j}\mapsto \mathsf{\bar{z}%
}_{ij}\bullet \mathsf{a}_{j}\in \mathsf{A}_{i}
\end{equation}%
is the expectation of observables at time $t_{j}$ conditional on observables
at time $t_{i}$. The strictly-positive state-price deflator $\mathsf{w}_{ij}$
in this expression rescales the measure of an event in the economic model to
the Arrow-Debreu price of the corresponding digital option. Consistency
among the family of price operations requires that their compositions
satisfy the tower law, a property that follows when the state-price
deflators take the form $\mathsf{w}_{ij}=\mathsf{w}_{i}/\mathsf{w}_{j}$ for
a strictly-positive numeraire process $\mathsf{w}$. The process $\mathsf{a}$
is then the price of a tradable security if it satisfies:%
\begin{equation}
\frac{\mathsf{a}_{i}}{\mathsf{w}_{i}}=\mathsf{\bar{z}}_{ij}\bullet \frac{%
\mathsf{a}_{j}}{\mathsf{w}_{j}}
\end{equation}%
for each time interval $t_{i}\leq t_{j}$. The numeraire has the dimension of
currency, so that the ratio of price over numeraire is dimensionless, and
the price model states that this ratio is a martingale in the economic
measure.

The price model decomposes into two components: the specification of an
economic model that quantifies the conditional expectations of economic
observables; and the identification of a numeraire process that adjusts
these expectations to match underlying prices. The price model is not
uniquely determined by the founding economic principles, and further
principles are needed to explain the origin of the numeraire. Completion of
the model requires an understanding of the roles that \emph{funding} and 
\emph{hedging} play in the optimisation of trading activity, leading to
additional guidelines that crystallise the price of the derivative.

\section{Funding}

The economic principles govern pricing for all market participants, but do
not fully determine price. Flexibility in the framework, exemplified by the
unidentified numeraire $\mathsf{w}$, is only resolved by looking more
closely at the activities within the financial institution.

Trading activity is funded by issuing bonds and shares, and the costs of
this operation are charged to the desk via the strictly-positive \emph{%
unsecured funding price} $\mathsf{u}$ representing the unit price of funding
for the institution. Benchmarking against the cost of funding allows
settlements at different times to be compared on a consistent basis, and
this makes the unsecured funding price the natural candidate for numeraire
in the price model. This hypothesis is overly restrictive, however, as there
is no facility in the approach to mark the model to market. The contribution
of funding to price is instead revealed by inspecting the funding
settlements alongside those from the traded security in the context of the
price model.

Consider the purchase at time $t$ for price $\mathsf{a}$ and the subsequent
sale at time $t+dt$ for price $\mathsf{a}+d\mathsf{a}$ of a tradable
security transacted over the finite time increment $dt$. The price model
relates these initial and terminal prices via the local martingale condition:%
\begin{equation}
0=\mathsf{\bar{z}}\bullet d\frac{\mathsf{a}}{\mathsf{w}}
\end{equation}%
for the conditional expectation from time $t+dt$ to time $t$ of the
dimensionless ratio $\mathsf{a}/\mathsf{w}$. The purchase of the security is
funded with the sale of $\mathsf{a}/\mathsf{u}$ units of unsecured funding,
returning at time $t+dt$ the proceeds $\mathsf{a}+d\mathsf{a}$ from the
resale of the security minus the costs $(\mathsf{a}/\mathsf{u})(\mathsf{u}+d%
\mathsf{u})$ from the repurchase of unsecured funding. Feeding these
settlements into the price model leads to the alternative local martingale
condition:%
\begin{equation}
0=\mathsf{\bar{z}}\bullet \frac{d\mathsf{a}-(\mathsf{a}/\mathsf{u})\,d%
\mathsf{u}}{\mathsf{w}+d\mathsf{w}}=\mathsf{\bar{z}}\bullet \frac{\mathsf{u}%
+d\mathsf{u}}{\mathsf{w}+d\mathsf{w}}\,d\frac{\mathsf{a}}{\mathsf{u}}
\end{equation}%
From the perspective of the financial institution, there are now two
representations of the local martingale condition for the same transaction.
Fortunately, these conditions are aligned when the dimensionless ratio $%
\mathsf{u}/\mathsf{w}$ is a martingale:%
\begin{equation}
0=\mathsf{\bar{z}}\bullet d\frac{\mathsf{u}}{\mathsf{w}}
\end{equation}%
The funding of trading activity thus imposes a normalisation constraint on
the model: the numeraire does not necessarily equal the unsecured funding
price, but their ratio is required to be driftless in the economic measure.

Recognising the elevated status of the unsecured funding price, the price
measure $\mathsf{\hat{z}}$ is defined to be the measure equivalent to the
economic measure $\mathsf{\bar{z}}$ with Radon-Nikodym kernel given by the
martingale $\mathsf{u}/\mathsf{w}$. The price model is expressed in terms of
this measure by the local martingale condition:%
\begin{equation}
0=\mathsf{\hat{z}}\bullet \frac{d\mathsf{a}-(\mathsf{a}/\mathsf{u})\,d%
\mathsf{u}}{\mathsf{u}+d\mathsf{u}}=\mathsf{\hat{z}}\bullet d\frac{\mathsf{a}%
}{\mathsf{u}}
\end{equation}%
for the price $\mathsf{a}$ of a tradable security. In this form, the
unsecured funding price $\mathsf{u}$ replaces the numeraire $\mathsf{w}$ as
discounting for the security. Further principles are then needed to identify
the price measure from the economic measure.

In addition to prescribing settlements, the derivative contract may also
stipulate terms for its funding. Defining the price of the derivative as the
discounted expectation in the price measure of its terminal settlements
neglects these incremental funding settlements. More accurately, the
derivative should be accounted for continuously, including any margin or
collateral payments, or any other costs incurred by the use of financial
resources.

The crucial consideration is the incremental settlement implied by the terms
of the contract, and in this there is commonality across different
derivative types. In each case, the derivative is associated with a \emph{%
market price} $\mathsf{a}$ and a strictly-positive \emph{funding price} $%
\mathsf{b}$ used to determine the net profit/loss over the time increment $%
dt $ for the self-funded position:%
\begin{equation}
d\mathsf{a}-\frac{\mathsf{a}}{\mathsf{b}}d\mathsf{b}=(\mathsf{b}+d\mathsf{b}%
)\,d\frac{\mathsf{a}}{\mathsf{b}}
\end{equation}%
This expression, which includes funding costs but neglects additional
charges for reserves against potential unwind costs, comprises the
profit/loss from derivative minus the profit/loss from funding, and the
owner receives these net settlements for as long as they hold the
derivative. Price is thus derived from the terminal settlements and binding
provisions for funding as specified in the contract, varying according to
the nature of the derivative.

\begin{figure}[p]
\centering
\includegraphics[width=0.95%
\linewidth]{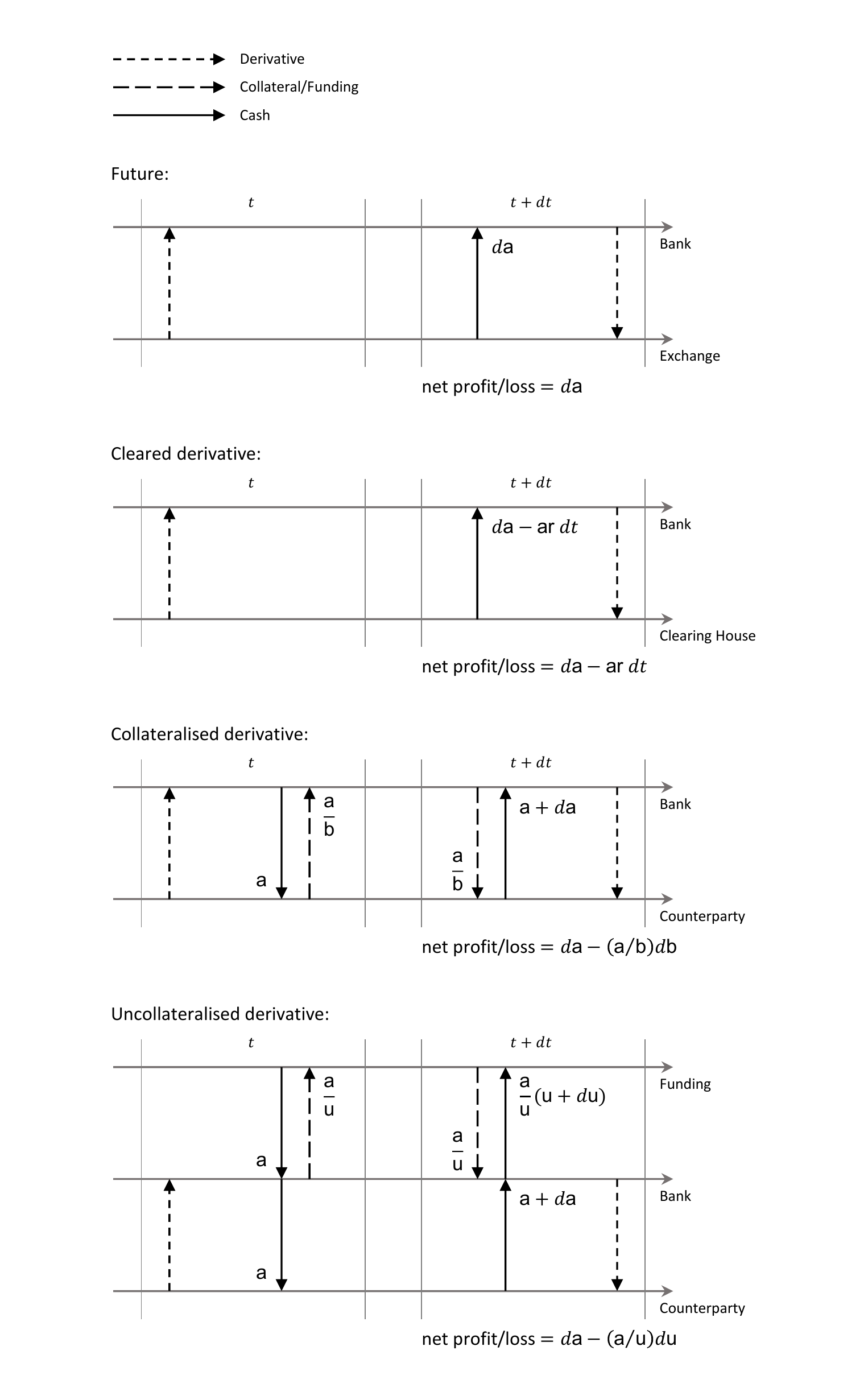}
\caption{The execution and settlement exchanges associated with four types
of derivative. Common to all these is the expression for the net profit/loss
over the time interval, equal to the price change minus the cost of funding
as prescribed by the derivative contract.}
\end{figure}

\begin{description}
\item[Futures] \negthinspace are standardised derivatives traded on
exchanges, typically with high liquidity and low transaction costs.
Participants use limit and market orders to submit prices and execute
trades, thereafter committing the holder to the series of payments required
to maintain the margin account. The net settlement over the time increment
is the variation margin:%
\begin{equation}
d\mathsf{a}
\end{equation}%
on the market price $\mathsf{a}$ of the future. In the general expression,
this is equivalent to funding with the unit $\mathsf{b}=1$.

\item[Cleared derivatives] \negthinspace extend the essential features of
futures to a wider range of standardised derivatives. Market makers provide
firm prices, and execution is intermediated by a clearing house whose
exposure to default risk is mitigated through the maintenance of an
interest-bearing margin account. The market price $\mathsf{a}$ serves as the
reference for the variation margin settlement:%
\begin{equation}
d\mathsf{a}-\mathsf{ar}\,dt
\end{equation}%
equivalent to funding with the cash account $\mathsf{b}=(1+\mathsf{r}%
\,dt)^{t/dt}$ that accrues interest at the rate $\mathsf{r}$ specified by
the clearing house.

\item[Collateralised derivatives] \negthinspace are bilateral transactions
commonly used for more illiquid structures. Default risk is mitigated
through the provision of collateral, whose terms are dictated by the legal
relationship between the\ two parties. Collateral is continuously rebalanced
to ensure it covers the derivative valuation. The net settlement over the
time increment is then:%
\begin{equation}
d\mathsf{a}-\frac{\mathsf{a}}{\mathsf{b}}d\mathsf{b}
\end{equation}%
being the variation of the derivative valuation $\mathsf{a}$ and the
collateral valuation $\mathsf{b}$ in the collateral account.

\item[Uncollateralised derivatives] \negthinspace offer bespoke derivative
features to clients without requiring the exchange of collateral. The cost
of unsecured funding, covering the anticipated closing price of the
derivative, is charged back to the desk. The net settlement in the trading
book is:%
\begin{equation}
d\mathsf{a}-\frac{\mathsf{a}}{\mathsf{u}}d\mathsf{u}
\end{equation}%
netting the variation on the derivative valuation $\mathsf{a}$ and the
incremental maintenance cost for unsecured funding with price $\mathsf{u}$.
\end{description}

For each of these types, holding the derivative becomes a commitment to a
series of self-funded net settlements in the trading book, entered at zero
cost and returning an amount that nets the variations on the market and
funding valuations for the derivative. Inserting these initial and terminal
prices into the price model leads to the local martingale condition:%
\begin{equation}
0=\mathsf{\hat{z}}\bullet \frac{d\mathsf{a}-(\mathsf{a}/\mathsf{b})\,d%
\mathsf{b}}{\mathsf{u}+d\mathsf{u}}=\mathsf{\hat{z}}\bullet \frac{\mathsf{b}%
+d\mathsf{b}}{\mathsf{u}+d\mathsf{u}}\,d\frac{\mathsf{a}}{\mathsf{b}}
\end{equation}%
for the conditional expectation from time $t+dt$ to time $t$ of the
dimensionless price ratio $\mathsf{a}/\mathsf{b}$. Margin settlements are
made at the endpoints of each interval in a series $t_{0}<\cdots <t_{n}$ of
times, and the local condition above is compounded to the martingale
condition:%
\begin{equation}
\frac{\mathsf{a}_{i}}{\mathsf{b}_{i}}=\mathsf{\hat{z}}_{ij}\bullet
\prod_{k=i}^{j-1}\frac{\mathsf{b}_{k+1}/\mathsf{u}_{k+1}}{\mathsf{\hat{z}}%
_{k\,k+1}\bullet \mathsf{b}_{k+1}/\mathsf{u}_{k+1}}\frac{\mathsf{a}_{j}}{%
\mathsf{b}_{j}}
\end{equation}%
The role of numeraire in this expression is taken by the funding price, and
the ratio of market price over funding price is a martingale in a measure
associated with the funding that is equivalent to the price measure.

Utilising the funding price as numeraire resolves the discounting question
for the settlements of the derivative. The correction to the price measure
is required to ensure consistency across the models generated for different
funding arrangements. In each case, the change of measure is driven by the
convexity between the funding price $\mathsf{b}$ of the derivative and the
unsecured funding price $\mathsf{u}$ of the financial institution, with
Radon-Nikodym kernel given by the dimensionless funding ratio $\mathsf{b}/%
\mathsf{u}$ drift-adjusted to be a martingale in the price measure.

A natural hierarchy exists in the market, with futures and cleared
derivatives providing the price transparency and liquidity to mark-to-market
bilateral derivatives. This market information is absorbed in the price
model through the mechanism of hedging, with the price measure emerging from
a martingale condition derived from fair pricing principles.

\section{Hedging}

Bridging the gap between the subjective expectations encapsulated in the
model for the economy and the market expectations expressed by the prices of
liquid securities is not without ambiguity, but guidance can be found in
traditional methods of portfolio optimisation. In the following, this
ambiguity is resolved by removing bias in optimal strategies based on the
mean and variance of portfolio returns, exploiting the market stratification
into liquid \emph{underlying securities} used for hedging and \emph{%
derivative securities} marked-to-market following fair pricing principles.

The performance of the strategy is measured relative to the unsecured
funding price $\mathsf{u}$ of the financial institution using the statistics
from the economic model $\mathsf{\bar{z}}$. For the strategy with price $%
\mathsf{a}$, performance is quantified by the mean and variance of the
increment $d(\mathsf{a}/\mathsf{u})$, delimiting the expected range for the
return on the strategy in proportion to the reference price. Benchmarking
against unsecured funding removes the dependency on the denominating
currency and allows the consistent evaluation of settlements at different
times.

While this is an effective method for combining economic and market
expectations in the price model, it is not necessarily an accurate
representation of risk management activities in a financial institution.
Both the economic model and the unsecured funding price depend on the
institution, and the resulting price model is sensitive to these components.
Furthermore, using mean and variance as targets for optimisation may not be
suited to the management of extreme events, and alternative statistics will
lead to variations in the derivative price. For these reasons, the principle
of portfolio optimisation that completes the price model is here held
separate from the three core economic principles.

\subsection{Portfolio optimisation}

The market provides access to a set of liquid underlying securities whose
vectors of market prices $\mathsf{p}$ and funding prices $\mathsf{q}$ are
directly observed. Normalised using the unsecured funding price $\mathsf{u}$%
, the return from the underlying securities over the trading interval is:%
\begin{equation}
\mathsf{R}_{\mathsf{p}}=\frac{d\mathsf{p}-(\mathsf{p}/\mathsf{q})\,d\mathsf{q%
}}{\mathsf{u}+d\mathsf{u}}=\frac{\mathsf{q}+d\mathsf{q}}{\mathsf{u}+d\mathsf{%
u}}\,d\frac{\mathsf{p}}{\mathsf{q}}
\end{equation}%
The return is adjusted to account for the convexity between underlying and
unsecured funding, captured in the dimensionless funding ratio $\mathsf{q}/%
\mathsf{u}$. The anticipated range for the return vector $\mathsf{R}_{%
\mathsf{p}}$ is quantified in terms of the mean vector $\mathsf{M}_{\mathsf{p%
}}$ and covariance matrix $\mathsf{V}_{\mathsf{p}}$:%
\begin{align}
\mathsf{M}_{\mathsf{p}}& =\mathsf{\bar{z}}\bullet \mathsf{R}_{\mathsf{p}} \\
\mathsf{V}_{\mathsf{p}}& =\mathsf{\bar{z}}\bullet \mathsf{R}_{\mathsf{p}%
}^{2}-(\mathsf{\bar{z}}\bullet \mathsf{R}_{\mathsf{p}})^{2}  \notag
\end{align}%
The effectiveness of the investment strategy then depends on its success in
identifying portfolios that minimise risk for a target expected return.

The expected performance of the underlying portfolio is measured by the mean 
$\mathsf{m}=\alpha \cdot \mathsf{M}_{\mathsf{p}}$ and variance $\mathsf{v}%
=\alpha \cdot \mathsf{V}_{\mathsf{p}}\alpha $ of its return, where the
composition of the portfolio is specified by the weights $\alpha $. The
economic model considers the return on portfolios with zero variance to be
guaranteed, and unless the mean is also zero this implies the existence of
an arbitrage. The technical requirement for the economic model to avoid
arbitrage is:%
\begin{equation}
\ker [\mathsf{V}_{\mathsf{p}}]\cdot \mathsf{M}_{\mathsf{p}}=\{0\}
\end{equation}%
For convenience, the stronger condition $\ker [\mathsf{V}_{\mathsf{p}%
}]=\{0\} $ is assumed in the following. In the more general case, the
optimal strategies developed here are valid only when the economic model
satisfies the no-arbitrage constraint, and are then determined only up to
the addition of a zero-variance portfolio.

Optimal portfolio weights, achieving the minimum possible variance for a
target mean, satisfy the stationarity condition:%
\begin{align}
0& =\delta (\alpha \cdot \mathsf{V}_{\mathsf{p}}\alpha +\lambda (\alpha
\cdot \mathsf{M}_{\mathsf{p}}-\mathsf{m})) \\
& =\delta \alpha \cdot (2\mathsf{V}_{\mathsf{p}}\alpha +\lambda \mathsf{M}_{%
\mathsf{p}})  \notag
\end{align}%
under variations $\delta \alpha $ of the portfolio $\alpha $, where the
Lagrange multiplier $\lambda $ maps to the risk appetite of the investor.
The optimal portfolio is thus:%
\begin{equation}
\alpha \propto \mathsf{V}_{\mathsf{p}}^{-1}\mathsf{M}_{\mathsf{p}}
\end{equation}%
The mean and variance achieved by this strategy lie on the parabola
specified by the relation:%
\begin{equation}
\frac{\mathsf{m}}{\sqrt{\mathsf{v}}}=\sqrt{\mathsf{V}_{\mathsf{p}}^{-1}%
\mathsf{M}_{\mathsf{p}}\cdot \mathsf{M}_{\mathsf{p}}}
\end{equation}%
where the expression on the right quantifies market sentiment on the
expected return that is acceptable for a unit of risk. The optimal
investment strategy minimises variance by diversifying risks across the
underlying securities, with the risk appetite determining the position of
the strategy on the efficient frontier parabola in mean-variance space.

\subsection{Derivative pricing}

Now consider a stratified market comprising the underlying securities with
market prices $\mathsf{p}$ and funding prices $\mathsf{q}$ and a derivative
security with market price $\mathsf{a}$ and funding price $\mathsf{b}$. The
underlying securities constitute the hedge market for the derivative
security, with the observed prices of the former used to mark-to-market the
latter via fair pricing principles.

The strategy is optimised against the joint moments of the underlying return 
$\mathsf{R}_{\mathsf{p}}$ and the normalised return $\mathsf{R}_{\mathsf{a}}$
on the derivative:%
\begin{equation}
\mathsf{R}_{\mathsf{a}}=\frac{d\mathsf{a}-(\mathsf{a}/\mathsf{b})\,d\mathsf{b%
}}{\mathsf{u}+d\mathsf{u}}=\frac{\mathsf{b}+d\mathsf{b}}{\mathsf{u}+d\mathsf{%
u}}\,d\frac{\mathsf{a}}{\mathsf{b}}
\end{equation}%
In addition to the mean vector $\mathsf{M}_{\mathsf{p}}$ and covariance
matrix $\mathsf{V}_{\mathsf{p}}$ for the underlyings, the performance of the
combined portfolio is quantified in terms of the mean $\mathsf{M}_{\mathsf{a}%
}$ and variance $\mathsf{V}_{\mathsf{a}}$ for the derivative:%
\begin{align}
\mathsf{M}_{\mathsf{a}}& =\mathsf{\bar{z}}\bullet \mathsf{R}_{\mathsf{a}} \\
\mathsf{V}_{\mathsf{a}}& =\mathsf{\bar{z}}\bullet \mathsf{R}_{\mathsf{a}%
}^{2}-(\mathsf{\bar{z}}\bullet \mathsf{R}_{\mathsf{a}})^{2}  \notag
\end{align}%
and the cross-covariance vector $\mathsf{C}_{\mathsf{pa}}$:%
\begin{equation}
\mathsf{C}_{\mathsf{pa}}=\mathsf{\bar{z}}\bullet \mathsf{R}_{\mathsf{p}}%
\mathsf{R}_{\mathsf{a}}-(\mathsf{\bar{z}}\bullet \mathsf{R}_{\mathsf{p}})(%
\mathsf{\bar{z}}\bullet \mathsf{R}_{\mathsf{a}})
\end{equation}%
Strategies for optimising the hedged derivative portfolio are determined
from these statistics.

The introduction of the derivative expands the opportunities for
diversification, leading to an incremental improvement in the expected
return for a unit of risk from the optimal strategy:%
\begin{equation}
\begin{bmatrix}
\mathsf{V}_{\mathsf{p}} & \mathsf{C}_{\mathsf{pa}} \\ 
\mathsf{C}_{\mathsf{pa}}^{t} & \mathsf{V}_{\mathsf{a}}%
\end{bmatrix}%
^{-1}%
\begin{bmatrix}
\mathsf{M}_{\mathsf{p}} \\ 
\mathsf{M}_{\mathsf{a}}%
\end{bmatrix}%
\cdot 
\begin{bmatrix}
\mathsf{M}_{\mathsf{p}} \\ 
\mathsf{M}_{\mathsf{a}}%
\end{bmatrix}%
=\mathsf{V}_{\mathsf{p}}^{-1}\mathsf{M}_{\mathsf{p}}\cdot \mathsf{M}_{%
\mathsf{p}}+\frac{(\mathsf{M}_{\mathsf{a}}-\mathsf{V}_{\mathsf{p}}^{-1}%
\mathsf{M}_{\mathsf{p}}\cdot \mathsf{C}_{\mathsf{pa}})^{2}}{\mathsf{V}_{%
\mathsf{a}}-\mathsf{V}_{\mathsf{p}}^{-1}\mathsf{C}_{\mathsf{pa}}\cdot 
\mathsf{C}_{\mathsf{pa}}}
\end{equation}%
Activity in the underlying market uncovers the performance target of
participants, and this suggests a principle for marking-to-market the fair
price of the derivative.

\begin{description}
\item[The Principle of Portfolio Optimisation:] The derivative security does
not increase the expected return for a unit of risk.
\end{description}

\noindent As can be seen from the expression above, this principle leads to
the following condition for the fair price of the derivative:%
\begin{equation}
\mathsf{M}_{\mathsf{a}}=\mathsf{V}_{\mathsf{p}}^{-1}\mathsf{M}_{\mathsf{p}%
}\cdot \mathsf{C}_{\mathsf{pa}}
\end{equation}%
Any other price for the derivative could be exploited in a strategy that
out-performs the underlying market, and so market efficiency drives the
price toward this equilibrium.

An alternative approach looks directly at the optimal portfolio that hedges
the derivative return. For hedge weights $\beta $, the return on the hedged
derivative has mean $\mathsf{m}$ and variance $\mathsf{v}$ given by:%
\begin{align}
\mathsf{m}& =\mathsf{M}_{\mathsf{a}}-\beta \cdot \mathsf{M}_{\mathsf{p}} \\
\mathsf{v}& =\mathsf{V}_{\mathsf{a}}-2\beta \cdot \mathsf{C}_{\mathsf{pa}%
}+\beta \cdot \mathsf{V}_{\mathsf{p}}\beta  \notag
\end{align}%
The optimal hedge weights, achieving the minimum possible variance, then
satisfy the stationarity condition:%
\begin{align}
0& =\delta (\mathsf{V}_{\mathsf{a}}-2\beta \cdot \mathsf{C}_{\mathsf{pa}%
}+\beta \cdot \mathsf{V}_{\mathsf{p}}\beta ) \\
& =2\delta \beta \cdot (\mathsf{V}_{\mathsf{p}}\beta -\mathsf{C}_{\mathsf{pa}%
})  \notag
\end{align}%
under variations $\delta \beta $ of the portfolio $\beta $. The hedge
portfolio is thus:%
\begin{equation}
\beta =\mathsf{V}_{\mathsf{p}}^{-1}\mathsf{C}_{\mathsf{pa}}
\end{equation}%
The mean and variance achieved by this strategy are:%
\begin{align}
\mathsf{m}& =\mathsf{M}_{\mathsf{a}}-\mathsf{V}_{\mathsf{p}}^{-1}\mathsf{M}_{%
\mathsf{p}}\cdot \mathsf{C}_{\mathsf{pa}} \\
\mathsf{v}& =\mathsf{V}_{\mathsf{a}}-\mathsf{V}_{\mathsf{p}}^{-1}\mathsf{C}_{%
\mathsf{pa}}\cdot \mathsf{C}_{\mathsf{pa}}  \notag
\end{align}%
The hedge does not eliminate the market risk of the derivative, but this
strategy reduces it to the minimum achievable. This suggests an alternative
principle for determining the fair price of the derivative.

\begin{description}
\item[The Principle of Portfolio Optimisation:] The return on a hedged
derivative security is unbiased.
\end{description}

\noindent The implied expression for the fair price of the derivative is
then:%
\begin{equation}
\mathsf{M}_{\mathsf{a}}=\mathsf{V}_{\mathsf{p}}^{-1}\mathsf{M}_{\mathsf{p}%
}\cdot \mathsf{C}_{\mathsf{pa}}
\end{equation}%
The efficient market equilibrates at the level that removes bias in the
residual hedged return, thereby calibrating the derivative price to the
market. Conveniently, both principles for marking-to-market the derivative
lead to the same fair pricing expression.

Market completeness -- the capacity of the market to replicate the
sensitivities of the derivative price -- impacts the residual variance of
the hedged derivative. Consider the underlying market with two sectors whose
returns are $\mathsf{R}_{\mathsf{p}}$ and $\mathsf{R}_{\mathsf{o}}$
respectively. The optimal hedge in this case comprises a portfolio $\beta _{%
\mathsf{p}}$ in the first sector and a portfolio $\beta _{\mathsf{o}}$ in
the second sector:%
\begin{align}
\begin{bmatrix}
\beta _{\mathsf{p}} \\ 
\beta _{\mathsf{o}}%
\end{bmatrix}%
=\,& 
\begin{bmatrix}
\mathsf{V}_{\mathsf{p}}^{-1}\mathsf{C}_{\mathsf{pa}} \\ 
0%
\end{bmatrix}
\\
& +%
\begin{bmatrix}
-\mathsf{V}_{\mathsf{p}}^{-1}\mathsf{C}_{\mathsf{po}}(\mathsf{V}_{\mathsf{o}%
}-\mathsf{C}_{\mathsf{po}}^{t}\mathsf{V}_{\mathsf{p}}^{-1}\mathsf{C}_{%
\mathsf{po}})^{-1}(\mathsf{C}_{\mathsf{oa}}-\mathsf{C}_{\mathsf{po}}^{t}%
\mathsf{V}_{\mathsf{p}}^{-1}\mathsf{C}_{\mathsf{pa}}) \\ 
(\mathsf{V}_{\mathsf{o}}-\mathsf{C}_{\mathsf{po}}^{t}\mathsf{V}_{\mathsf{p}%
}^{-1}\mathsf{C}_{\mathsf{po}})^{-1}(\mathsf{C}_{\mathsf{oa}}-\mathsf{C}_{%
\mathsf{po}}^{t}\mathsf{V}_{\mathsf{p}}^{-1}\mathsf{C}_{\mathsf{pa}})%
\end{bmatrix}
\notag
\end{align}%
The first term on the right is the optimal strategy when only the first
sector is available for hedging. The second term improves hedge performance
by substituting part of the first sector hedge with securities from the
second sector that better replicate the return on the derivative. This leads
to an incremental reduction in the variance of the hedged derivative:%
\begin{align}
\mathsf{v}=\,& \mathsf{V}_{\mathsf{a}}-\mathsf{V}_{\mathsf{p}}^{-1}\mathsf{C}%
_{\mathsf{pa}}\cdot \mathsf{C}_{\mathsf{pa}} \\
& -(\mathsf{V}_{\mathsf{o}}-\mathsf{C}_{\mathsf{po}}^{t}\mathsf{V}_{\mathsf{p%
}}^{-1}\mathsf{C}_{\mathsf{po}})^{-1}(\mathsf{C}_{\mathsf{oa}}-\mathsf{C}_{%
\mathsf{po}}^{t}\mathsf{V}_{\mathsf{p}}^{-1}\mathsf{C}_{\mathsf{pa}})\cdot (%
\mathsf{C}_{\mathsf{oa}}-\mathsf{C}_{\mathsf{po}}^{t}\mathsf{V}_{\mathsf{p}%
}^{-1}\mathsf{C}_{\mathsf{pa}})  \notag
\end{align}%
Each expansion in the range of hedge securities enhances the opportunities
for offsetting the risk of the derivative.

By setting the expected return on the hedged derivative to zero, the price
model interpolates between two accounting methodologies. At one extreme,
when the underlying market is empty price is identified with expectation. At
the opposite extreme, when the underlying market is complete price is
determined by replication. The extent to which the price model relies on
expectation versus replication depends on market completeness: the
sensitivity of price to the economic model is mitigated by the calibration
to available market prices; conversely, the economic model plugs the gap in
pricing when liquidity is limited.

\subsection{Hedge performance}

For the observables within a time interval, the economic model provides two
measures: the \emph{forecast} measure and the \emph{empirical} measure.

\begin{description}
\item[Forecast measure:] The measure $\mathsf{\bar{z}}^{s}$ seen at the
start of the time interval, capturing the predicted statistics for the
observables.

\item[Empirical measure:] The measure $\mathsf{\bar{z}}^{e}$ seen at the end
of the time interval, capturing the observed statistics for the observables.
\end{description}

\noindent Construction of the economic model begins with a reduction to the
macroscopic variables that describe the economy, and proceeds with the
consistent assignment of probabilities for the values of these observables.
Evidence refines these assignments: the measure starts as informed guess and
ends as statistical analysis. Performance of the economic model is then
quantified by the gap between its forecast and empirical measures.

These measures respectively generate the hedge strategies:%
\begin{align}
\beta ^{s}& =\mathsf{V}_{\mathsf{p}}^{s}{}^{-1}\mathsf{C}_{\mathsf{pa}}^{s}
\\
\beta ^{e}& =\mathsf{V}_{\mathsf{p}}^{e}{}^{-1}\mathsf{C}_{\mathsf{pa}}^{e} 
\notag
\end{align}%
The first portfolio is the optimal strategy computed at the start of the
interval, based on initial expectations for the future of the economy. The
second portfolio is the optimal strategy computed at the end of the
interval, hedging with the benefit of hindsight according to the realised
market volatility over the interval.

The executed hedge strategy is necessarily derived from the forecast
measure. Hedge performance is then quantified by the empirical mean $\mathsf{%
m}$ and variance $\mathsf{v}$ of the return on the hedged derivative:%
\begin{align}
\mathsf{m}& =(\mathsf{M}_{\mathsf{a}}^{e}-\mathsf{V}_{\mathsf{p}}^{e}{}^{-1}%
\mathsf{M}_{\mathsf{p}}^{e}\cdot \mathsf{C}_{\mathsf{pa}}^{e})+(\beta
^{e}-\beta ^{s})\cdot \mathsf{M}_{\mathsf{p}}^{e} \\
\mathsf{v}& =(\mathsf{V}_{\mathsf{a}}^{e}-\mathsf{V}_{\mathsf{p}}^{e}{}^{-1}%
\mathsf{C}_{\mathsf{pa}}^{e}\cdot \mathsf{C}_{\mathsf{pa}}^{e})+(\beta
^{e}-\beta ^{s})\cdot \mathsf{V}_{\mathsf{p}}^{e}(\beta ^{e}-\beta ^{s}) 
\notag
\end{align}%
The empirical variance decomposes into two components. The market risk
component is the minimum variance achievable in the empirical measure using
the empirical hedge strategy. The model risk component is the additional
variance due to the sub-optimality of the forecast hedge strategy.

\begin{figure}[!t]
\centering
\includegraphics[width=0.95%
\linewidth]{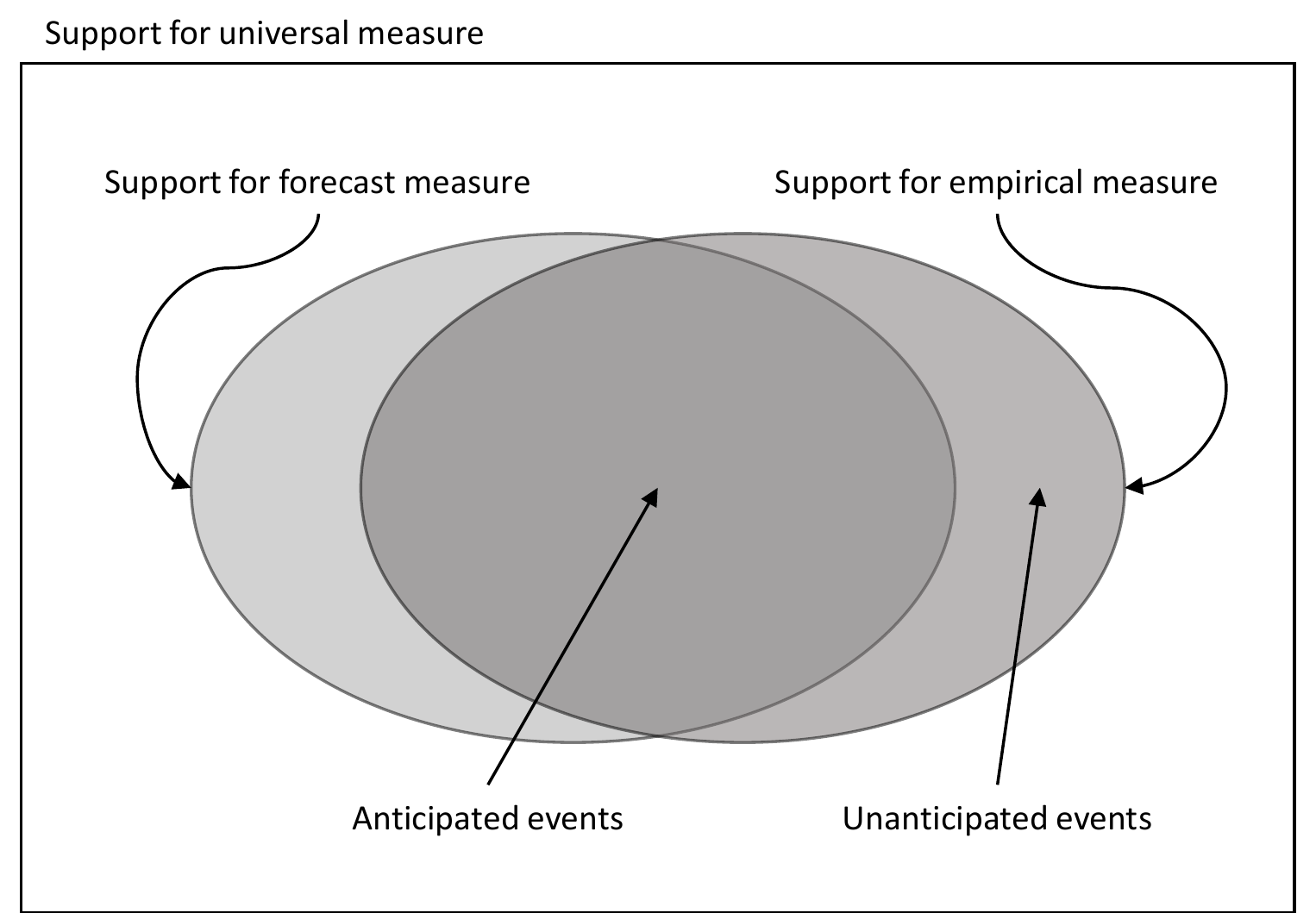}
\caption{The support for the empirical measure divides into the anticipated
events shared with the forecast measure and the unanticipated events with
zero forecast measure.}
\end{figure}

The forecast and empirical measures are not necessarily equivalent: events
that are anticipated may not materialise; conversely, and more dangerously,
unanticipated `black swan' events accounted in the empirical measure are
neglected in the forecast measure. Introduce a universal measure $\mathsf{z}$
that supports the events for both measures, with Radon-Nikodym kernels
expressed as:%
\begin{align}
\frac{d\mathsf{\bar{z}}^{s}}{d\mathsf{z}}& =\mathsf{A} \\
\frac{d\mathsf{\bar{z}}^{e}}{d\mathsf{z}}& =\mathsf{A}+\varepsilon \mathsf{D}
\notag
\end{align}%
The empirical measure corrects the forecast measure via the term $%
\varepsilon \mathsf{D}$. Considering the case when $\varepsilon $ is small,
the difference between the empirical and forecast hedges is expanded:%
\begin{equation}
\beta ^{e}-\beta ^{s}=\mathsf{V}_{\mathsf{p}}^{s}{}^{-1}\Omega
+O[\varepsilon ^{2}]
\end{equation}%
where:%
\begin{equation}
\Omega =\mathsf{z}\bullet \varepsilon \mathsf{D}(\mathsf{R}_{\mathsf{p}}-%
\mathsf{\bar{z}}^{s}\bullet \mathsf{R}_{\mathsf{p}})(\mathsf{R}_{\mathsf{a}%
}-\beta ^{s}\cdot \mathsf{R}_{\mathsf{p}})
\end{equation}%
Sub-optimality of the forecast hedge is proportional to the deviation of the
empirical measure from the forecast measure. It is also proportional to the
residual return from the hedged derivative, confirming that market
completeness is a determinant for the performance of the hedge.

Anticipated and unanticipated events are separated in the correction term:%
\begin{equation}
\mathsf{D}=\mathsf{SA}+\mathsf{B}
\end{equation}%
where $\mathsf{B}$ satisfies $\mathsf{AB}=0$. Relative to the universal
measure, the empirical measure deviates from the forecast measure by the
kernel $\varepsilon (\mathsf{SA}+\mathsf{B})$ where $\varepsilon \mathsf{SA}$
corrects the measure for anticipated events and $\varepsilon \mathsf{B}$
appends the measure for unanticipated events. The expression for the hedge
error decomposes as:%
\begin{equation}
\Omega =\Omega _{a}+\Omega _{b}
\end{equation}%
The first term in the decomposition:%
\begin{equation}
\Omega _{a}=\mathsf{\bar{z}}^{s}\bullet \varepsilon \mathsf{S}(\mathsf{R}_{%
\mathsf{p}}-\mathsf{\bar{z}}^{s}\bullet \mathsf{R}_{\mathsf{p}})(\mathsf{R}_{%
\mathsf{a}}-\beta ^{s}\cdot \mathsf{R}_{\mathsf{p}})
\end{equation}%
is the hedge error arising from the adjustment $\mathsf{S}$ that rescales
the probabilities of anticipated events following the assessment of actual
market volatility over the interval. The second term in the decomposition:%
\begin{equation}
\Omega _{b}=\mathsf{\bar{z}}^{e}\bullet (\mathsf{B}\neq 0)(\mathsf{R}_{%
\mathsf{p}}-\mathsf{\bar{z}}^{s}\bullet \mathsf{R}_{\mathsf{p}})(\mathsf{R}_{%
\mathsf{a}}-\beta ^{s}\cdot \mathsf{R}_{\mathsf{p}})
\end{equation}%
is the hedge error arising from the unanticipated events indicated by the
condition $\mathsf{B}\neq 0$. The first term, the `known-unknown'
contribution to model risk, is evaluated within the context of the forecast
measure by estimating the error bounds for model parameters. The second
term, the `unknown-unknown' contribution to model risk, is harder to assess
in advance as it is measured against events that are initially assumed to be
impossible.

\section{Expectation and price}

The founding economic principles dictate that the incremental change in the
price ratio of the derivative security satisfies the local martingale
condition:%
\begin{equation}
0=\mathsf{\hat{z}}\bullet \frac{\mathsf{b}+d\mathsf{b}}{\mathsf{u}+d\mathsf{u%
}}\,d\frac{\mathsf{a}}{\mathsf{b}}
\end{equation}%
where the price measure $\mathsf{\hat{z}}$ adjusts the economic measure $%
\mathsf{\bar{z}}$, calibrating to the underlying securities via the local
martingale condition:%
\begin{equation}
0=\mathsf{\hat{z}}\bullet \frac{\mathsf{q}+d\mathsf{q}}{\mathsf{u}+d\mathsf{u%
}}\,d\frac{\mathsf{p}}{\mathsf{q}}
\end{equation}%
for the incremental changes in the price ratios of the underlying
securities. This price model satisfies the principles of replication and
economic equivalence by construction; it satisfies the principle of
no-arbitrage when the price measure is positive.

While the role of economic expectation in price determination is clear from
these local martingale conditions, the economic principles and market
calibrations do not uniquely determine the price measure. Taking its steer
from portfolio optimisation, the previous section derives an additional
condition for the fair price of the derivative by removing bias in the
return from the hedged portfolio. This resolves the ambiguity in the price
model.

Expanding the terms in the fair pricing condition that include the return on
the derivative security leads to:%
\begin{equation}
\mathsf{\bar{z}}\bullet \mathsf{R}_{\mathsf{a}}=\mathsf{\bar{z}}\bullet 
\mathsf{V}_{\mathsf{p}}^{-1}\mathsf{M}_{\mathsf{p}}\cdot (\mathsf{R}_{%
\mathsf{p}}-\mathsf{\bar{z}}\bullet \mathsf{R}_{\mathsf{p}})\mathsf{R}_{%
\mathsf{a}}
\end{equation}%
This is re-arranged to the following relation for the incremental change in
the derivative price ratio:%
\begin{equation}
0=\mathsf{\bar{z}}[\alpha ]\bullet \frac{\mathsf{b}+d\mathsf{b}}{\mathsf{u}+d%
\mathsf{u}}\,d\frac{\mathsf{a}}{\mathsf{b}}
\end{equation}%
where the dimensionless parameters $\alpha $ are marked-to-market via the
relation:%
\begin{equation}
0=\mathsf{\bar{z}}[\alpha ]\bullet \frac{\mathsf{q}+d\mathsf{q}}{\mathsf{u}+d%
\mathsf{u}}\,d\frac{\mathsf{p}}{\mathsf{q}}
\end{equation}%
for the incremental changes in the underlying price ratios. The price
measure $\mathsf{\hat{z}}=\mathsf{\bar{z}}[\alpha ]$ in these expressions is
equivalent to the economic measure $\mathsf{\bar{z}}$, with Radon-Nikodym
kernel adjusting for the returns on the underlying securities:%
\begin{equation}
\frac{d\mathsf{\bar{z}}[\alpha ]}{d\mathsf{\bar{z}}}=1-\alpha \cdot (\frac{%
\mathsf{q}+d\mathsf{q}}{\mathsf{u}+d\mathsf{u}}\,d\frac{\mathsf{p}}{\mathsf{q%
}}-\mathsf{\bar{z}}\bullet \frac{\mathsf{q}+d\mathsf{q}}{\mathsf{u}+d\mathsf{%
u}}\,d\frac{\mathsf{p}}{\mathsf{q}})
\end{equation}%
In this representation, the funding price assumes the mantle of numeraire,
and the economic measure is tweaked to calibrate the price measure to market
prices. Direct comparison with the general expression confirms that this
price model satisfies the principles of replication and economic
equivalence. It does not, however, satisfy the principle of no-arbitrage: if
the underlying price changes deviate significantly from their expectations,
the kernel could be negative. This highlights the inadequacies of
mean-variance optimisation as a method of price determination. By
under-estimating the impact of tail events, these performance metrics leave
open the possibility of arbitrage.

The defect in the price model is resolved by substituting the linear kernel
with the following exponential alternative:%
\begin{equation}
\frac{d\mathsf{\bar{z}}[\alpha ]}{d\mathsf{\bar{z}}}=\frac{\exp \!\left[
-\alpha \cdot \dfrac{\mathsf{q}+d\mathsf{q}}{\mathsf{u}+d\mathsf{u}}\,d%
\dfrac{\mathsf{p}}{\mathsf{q}}\right] }{\mathsf{\bar{z}}\bullet \exp \!\left[
-\alpha \cdot \dfrac{\mathsf{q}+d\mathsf{q}}{\mathsf{u}+d\mathsf{u}}\,d%
\dfrac{\mathsf{p}}{\mathsf{q}}\right] }
\end{equation}%
This variant of the price model is justified as it converges to the original
when the underlying return is small and avoids arbitrage when the underlying
return is large, reweighting the performance measure to avoid arbitrage from
tail events. The market calibration is solved by iterating the
Newton-Raphson scheme:%
\begin{equation}
\alpha \mapsto \alpha +\mathsf{V}_{\mathsf{p}}[\alpha ]^{-1}\mathsf{M}_{%
\mathsf{p}}[\alpha ]
\end{equation}%
where the mean vector $\mathsf{M}_{\mathsf{p}}[\alpha ]$ and covariance
matrix $\mathsf{V}_{\mathsf{p}}[\alpha ]$ are computed using the equivalent
measure $\mathsf{\bar{z}}[\alpha ]$ from the preceding iteration. Starting
with $\alpha =0$, convergence of this scheme then relies on the
invertibility of the covariance at each iteration. Calibration determines a
portfolio of underlying securities whose return is exponentiated to generate
the kernel for the price measure relative to the economic measure, with
weights $\alpha $ that control the deviation of the underlying prices from
their expectations.

The price expression is compounded on the margin settlement schedule to
generate the martingale condition for the derivative price ratio:%
\begin{equation}
\frac{\mathsf{a}_{i}}{\mathsf{b}_{i}}=\mathsf{\bar{z}}_{ij}\bullet
\prod_{k=i}^{j-1}\frac{\exp [-\alpha _{k}\cdot (\mathsf{q}_{k+1}/\mathsf{u}%
_{k+1})\,d(\mathsf{p}_{k}/\mathsf{q}_{k})]\,\mathsf{b}_{k+1}/\mathsf{u}_{k+1}%
}{\mathsf{\bar{z}}_{k\,k+1}\bullet \exp [-\alpha _{k}\cdot (\mathsf{q}_{k+1}/%
\mathsf{u}_{k+1})\,d(\mathsf{p}_{k}/\mathsf{q}_{k})]\,\mathsf{b}_{k+1}/%
\mathsf{u}_{k+1}}\frac{\mathsf{a}_{j}}{\mathsf{b}_{j}}
\end{equation}%
This expression determines the price of the derivative as the expectation of
its discounted terminal settlement, where discounting matches funding and
the measure is adjusted to calibrate underlying prices and accommodate
convexity between the funding ratio and the returns on underlying securities.

\begin{figure}[!t]
\makebox[\textwidth][c]{\ 
\Ovalbox{
\begin{tabular}{c}
\\
\parbox{\textwidth}{\centering\large\textsf{\textbf{MAXIMUM ENTROPY PRICE MODEL}}}\\
\\
\begin{minipage}{\textwidth}\setlength{\parindent}{15pt}
\noindent The ingredients of the maximum entropy price model are the economic measure $\mathsf{\bar{z}}$ quantifying the expectations of economic observables, the unsecured funding price $\mathsf{u}$ of the financial institution, and the observed market prices $\mathsf{p}$ and funding prices $\mathsf{q}$ of liquid underlying securities.

For a derivative security with market price $\mathsf{a}$ and funding price $\mathsf{b}$, the price model is given by the local martingale condition:
\begin{equation*}
0=\mathsf{\bar{z}}\bullet \exp\!\left[-\alpha\cdot\frac{\mathsf{q}+d\mathsf{q}}{\mathsf{u}+d\mathsf{u}}\,d\dfrac{\mathsf{p}}{\mathsf{q}}\right]\frac{\mathsf{b}+d\mathsf{b}}{\mathsf{u}+d\mathsf{u}}\,d\dfrac{\mathsf{a}}{\mathsf{b}}
\end{equation*}
for the change in the derivative price ratio $\mathsf{a}/\mathsf{b}$ over the time step $dt$, where the coefficients $\alpha$\ mark the model to the underlying prices via the local martingale condition:
\begin{equation*}
0=\mathsf{\bar{z}}\bullet \exp\!\left[-\alpha\cdot\frac{\mathsf{q}+d\mathsf{q}}{\mathsf{u}+d\mathsf{u}}\,d\dfrac{\mathsf{p}}{\mathsf{q}}\right]\frac{\mathsf{q}+d\mathsf{q}}{\mathsf{u}+d\mathsf{u}}\,d\dfrac{\mathsf{p}}{\mathsf{q}}
\end{equation*}
for the change in the underlying price ratios $\mathsf{p}/\mathsf{q}$ over the time step $dt$.

The price model assumes the expected return on the funded derivative, normalised by the unsecured funding price, is zero in the price measure that calibrates the economic measure to underlying prices with minimum relative entropy.
\end{minipage}\\
\smallskip
\end{tabular}
}}
\end{figure}

In a Bayesian interpretation of the approach, the economic measure is
considered to be the maximum entropy state representing the best assumptions
on price in the absence of market information. Assessed relative to this
state, the price measure with exponential form for the kernel is the measure
that maximises entropy while calibrating to the available prices of
underlying securities. The price model, originally derived from the
principles of no-arbitrage and portfolio optimisation, is then equivalently
derived on the assumption of maximum entropy.

\begin{description}
\item[The Maximum Entropy Principle:] The price measure minimises entropy
relative to the economic measure subject to calibration to the prices of
liquid underlying securities.
\end{description}

\noindent Market intelligence on price is incomplete, and this principle
proposes that the information vacuum is filled by the expectations
quantified in the economic model. This is achieved by minimising the
relative entropy of the price measure from the economic measure, subject to
the mark-to-market constraints. The stationarity condition, whose solution
is the exponential kernel $d\mathsf{\bar{z}}[\alpha ]/d\mathsf{\bar{z}}$
defined above, is:%
\begin{align}
0& =\delta (\mathsf{\bar{z}}\bullet \mathsf{W}\log [\mathsf{W}]+\lambda (%
\mathsf{\bar{z}}\bullet \mathsf{W}-1)+\alpha \cdot \mathsf{\bar{z}}\bullet 
\mathsf{W}\frac{\mathsf{q}+d\mathsf{q}}{\mathsf{u}+d\mathsf{u}}\,d\frac{%
\mathsf{p}}{\mathsf{q}}) \\
& =\mathsf{\bar{z}}\bullet \delta \mathsf{W}(\log [\mathsf{W}]+(1+\lambda
)+\alpha \cdot \frac{\mathsf{q}+d\mathsf{q}}{\mathsf{u}+d\mathsf{u}}\,d\frac{%
\mathsf{p}}{\mathsf{q}})  \notag
\end{align}%
for variations $\delta \mathsf{W}$ of the kernel $\mathsf{W}$. The Lagrange
multipliers $\lambda $ and $\alpha $ are solved for the normalisation
constraint $\mathsf{\bar{z}}\bullet \mathsf{W}=1$ and the market
calibrations.

Maximum entropy states arise in thermodynamic applications for the
macroscopic variables of a system whose microscopic variables evolve on a
much shorter timescale. In this perspective, the portfolio weights $\alpha $
are the inverse-temperatures for the underlying securities -- `hot'
securities discover the equilibrium of the economic model, while `cold'
securities have prices that deviate from expectations. This analogy between
non-equilibrium thermodynamics and economic agents discovering price through
trading activity suggests an alternative route to the price model which will
not be pursued further here.

\section{Continuous settlement}

The incremental price condition is compounded on the margin settlement
schedule, and the ratio of market price over funding price for the
derivative is a martingale on this schedule in an equivalent measure that
calibrates the economic measure to underlying returns in each settlement
period. If margin is settled frequently, this can be approximated by a
continuous-time model.

\begin{figure}[p]
\centering
\includegraphics[width=0.95%
\linewidth]{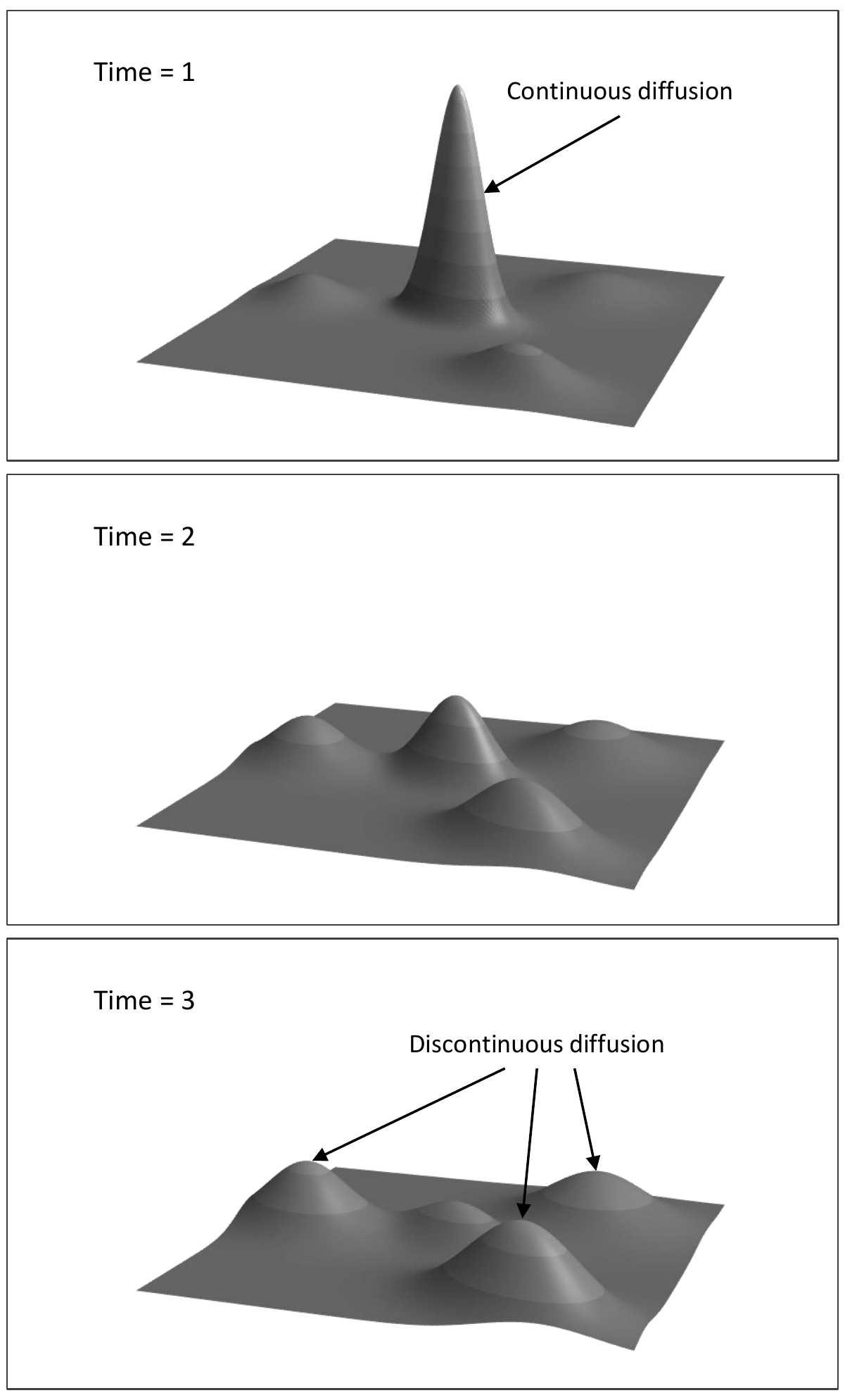}
\caption{The evolution of the continuous-time process has a continuous
component described by its mean and covariance rates and a discontinuous
component described by the frequency of jumps of different sizes.}
\end{figure}

The technical conditions that enable the perpetual subdivision of a process
are captured in the L\'{e}vy-Khintchine representation of its stochastic
differential equation. The evolution of the continuous-time vector process $%
\mathsf{s}$ in the measure $\mathsf{z}$ is described by its volatility
parameters $(\mu ,\nu ,\phi )$, where $\mu $ and $\nu $ are the rates of
change for the mean and covariance in the continuous diffusion, and $\phi $
is the frequency density for jumps in the discontinuous diffusion. The L\'{e}%
vy-Khintchine representation of the stochastic differential equation is then:%
\begin{align}
& \frac{\log \mathsf{z}\bullet \exp [k\cdot d\mathsf{s}]}{dt}= \\
& \hspace{1cm}k\cdot \mu +\frac{1}{2}k\cdot \nu k+\int (\exp [k\cdot
j]-1)\phi \lbrack j]\,dj+O[dt]  \notag
\end{align}%
The continuous diffusion captures the normal conditions of the evolution,
whose behaviour is accurately described over short intervals by the mean and
covariance in a Gauss distribution for the increment. The discontinuous
diffusion adjusts the skew and kurtosis and other higher moments of the
incremental distribution, and enables the modelling of regime switches where
the normal correlations between state variables are inverted.

Differentiating the stochastic differential equation repeatedly with respect
to the conjugate variable $k$ brings down a power of the increment $d\mathsf{%
s}$ in the integrand. This observation is used to identify the stochastic
differential equation for the derived process $f[t,\mathsf{s}]$ as a
function of time and the state process:%
\begin{align}
& \frac{\log \mathsf{z}\bullet \exp [k\,df[t,\mathsf{s}]]}{dt}= \\
& \hspace{1cm}k(\dot{f}[t,\mathsf{s}]+f^{\prime }[t,\mathsf{s}]\cdot \mu +%
\frac{1}{2}\func{tr}[f^{\prime \prime }[t,\mathsf{s}]\nu ])+\frac{1}{2}%
k^{2}f^{\prime }[t,\mathsf{s}]\cdot \nu f^{\prime }[t,\mathsf{s}]  \notag \\
& \hspace{1cm}+\int (\exp [k(f[t,\mathsf{s}+j]-f[t,\mathsf{s}])]-1)\phi
\lbrack j]\,dj+O[dt]  \notag
\end{align}%
This is the It\^{o} lemma in the L\'{e}vy-Khintchine representation.
Applications of the lemma transform the analysis of continuous-time
processes into the realm of partial differential-integral equations, and are
used in the following to derive hedge strategies and fair pricing conditions
for continuously-settled derivatives.

\subsection{Market and funding convexity}

The price model is based on the increments $d(\mathsf{p}/\mathsf{q})$ and $d(%
\mathsf{a}/\mathsf{b})$ for the price ratios of underlying and derivative
securities and the increments $d(\mathsf{q}/\mathsf{u})$ and $d(\mathsf{b}/%
\mathsf{u})$ for the ratios of the corresponding funding prices over the
unsecured funding price. These increments are modelled in the economic
measure $\mathsf{\bar{z}}$ by their combined stochastic differential
equation in the L\'{e}vy-Khintchine representation, enabling the
construction of optimal hedge strategies and the fair pricing condition from
the volatility parameters. The model then captures the \emph{market convexity%
} arising from nonlinearity in the relationships between market prices and
the \emph{funding convexity} resulting from differences in funding
arrangements.

To reduce the dimensionality of the problem, in this section it is assumed
that the underlying and derivative securities have the same funding price:%
\begin{equation}
\mathsf{q}=\mathsf{b}=\mathsf{fu}
\end{equation}%
scaling the unsecured funding price by the strictly-positive funding ratio $%
\mathsf{f}$. This simplification puts the focus primarily on market
convexity. The neglected convexities between underlying and derivative
funding can be accommodated in the framework by extending the volatility
parameters to allow decorrelation between the funding prices.

The joint evolution for the funding ratio $\mathsf{f}$ and the price ratios $%
\mathsf{s}=\mathsf{p}/\mathsf{q}$ and $\mathsf{c}=\mathsf{a}/\mathsf{b}$ of
underlying and derivative securities is modelled in the economic measure $%
\mathsf{\bar{z}}$ by the stochastic differential equation:%
\begin{align}
& \frac{\log \mathsf{\bar{z}}\bullet \exp [k_{\mathsf{f}}\,d\mathsf{f}/%
\mathsf{f}+k_{\mathsf{s}}\cdot d\mathsf{s}+k_{\mathsf{c}}\,d\mathsf{c}]}{dt}=
\\
& \hspace{1cm}%
\begin{bmatrix}
k_{\mathsf{f}} \\ 
k_{\mathsf{s}} \\ 
k_{\mathsf{c}}%
\end{bmatrix}%
\cdot 
\begin{bmatrix}
\mu _{\mathsf{f}} \\ 
\mu _{\mathsf{s}} \\ 
\mu _{\mathsf{c}}%
\end{bmatrix}%
+\frac{1}{2}%
\begin{bmatrix}
k_{\mathsf{f}} \\ 
k_{\mathsf{s}} \\ 
k_{\mathsf{c}}%
\end{bmatrix}%
\cdot 
\begin{bmatrix}
\nu _{\mathsf{f}} & \nu _{\mathsf{fs}}^{t} & \nu _{\mathsf{fc}} \\ 
\nu _{\mathsf{fs}} & \nu _{\mathsf{s}} & \nu _{\mathsf{sc}} \\ 
\nu _{\mathsf{fc}} & \nu _{\mathsf{sc}}^{t} & \nu _{\mathsf{c}}%
\end{bmatrix}%
\begin{bmatrix}
k_{\mathsf{f}} \\ 
k_{\mathsf{s}} \\ 
k_{\mathsf{c}}%
\end{bmatrix}
\notag \\
& \hspace{1cm}+\int (\exp [k_{\mathsf{f}}j_{\mathsf{f}}+k_{\mathsf{s}}\cdot
j_{\mathsf{s}}+k_{\mathsf{c}}j_{\mathsf{c}}]-1)\phi \lbrack j_{\mathsf{f}%
},j_{\mathsf{s}},j_{\mathsf{c}}]\,dj_{\mathsf{f}}\,dj_{\mathsf{s}}\,dj_{%
\mathsf{c}}+O[dt]  \notag
\end{align}%
The continuous diffusion is described by the mean rates $\mu _{\mathsf{f}}$, 
$\mu _{\mathsf{s}}$ and $\mu _{\mathsf{c}}$ and the covariance rates $\nu _{%
\mathsf{f}}$, $\nu _{\mathsf{s}}$ and $\nu _{\mathsf{c}}$ of the funding and
price ratios, together with the cross-covariance rates $\nu _{\mathsf{fs}}$, 
$\nu _{\mathsf{fc}}$ and $\nu _{\mathsf{sc}}$ between them. The
discontinuous diffusion is described by the frequency density $\phi \lbrack
j_{\mathsf{f}},j_{\mathsf{s}},j_{\mathsf{c}}]$ for jumps $j_{\mathsf{f}}$, $%
j_{\mathsf{s}}$ and $j_{\mathsf{c}}$ in the funding and price ratios.

The model is used to generate statistics in the equivalent measure $\mathsf{%
\bar{z}}[\alpha ]$, where the parameters $\alpha $ that adjust the economic
measure will be used to mark the model to market. The returns $\mathsf{R}_{%
\mathsf{p}}$ and $\mathsf{R}_{\mathsf{a}}$ are expressed in terms of the
increments $d\mathsf{f}/\mathsf{f}$, $d\mathsf{s}$ and $d\mathsf{c}$:%
\begin{equation}
\mathsf{R}_{\mathsf{p}}=\mathsf{f}(1+\frac{d\mathsf{f}}{\mathsf{f}})\,d%
\mathsf{s\hspace{1cm}R}_{\mathsf{a}}=\mathsf{f}(1+\frac{d\mathsf{f}}{\mathsf{%
f}})\,d\mathsf{c}
\end{equation}%
The statistics for the underlying and derivative are encapsulated in the
stochastic differential equation for the returns as the coefficients in its
series expansion with respect to the conjugate variables. The It\^{o} lemma
derives the expression:%
\begin{align}
& \frac{\log \mathsf{\bar{z}}[\alpha ]\bullet \exp [k_{\mathsf{p}}\cdot 
\mathsf{R}_{\mathsf{p}}+k_{\mathsf{a}}\mathsf{R}_{\mathsf{a}}]}{dt}= \\
& \hspace{1cm}\mathsf{f}(k_{\mathsf{p}}\cdot (\mu _{\mathsf{s}}[\alpha ]+\nu
_{\mathsf{fs}})+k_{\mathsf{a}}(\mu _{\mathsf{c}}[\alpha ]+\nu _{\mathsf{fc}%
}))  \notag \\
& \hspace{1cm}+\mathsf{f}^{2}(\frac{1}{2}k_{\mathsf{p}}\cdot \nu _{\mathsf{s}%
}k_{\mathsf{p}}+\frac{1}{2}k_{\mathsf{a}}^{2}\nu _{\mathsf{c}}+k_{\mathsf{a}%
}k_{\mathsf{p}}\cdot \nu _{\mathsf{sc}})  \notag \\
& \hspace{1cm}+\int (\exp [\mathsf{f}(1+j_{\mathsf{f}})(k_{\mathsf{p}}\cdot
j_{\mathsf{s}}+k_{\mathsf{a}}j_{\mathsf{c}})]-1)\phi \lbrack \alpha ][j_{%
\mathsf{f}},j_{\mathsf{s}},j_{\mathsf{c}}]\,dj_{\mathsf{f}}\,dj_{\mathsf{s}%
}\,dj_{\mathsf{c}}+O[dt]  \notag
\end{align}%
with adjusted volatility parameters:%
\begin{align}
\mu _{\mathsf{s}}[\alpha ]& =\mu _{\mathsf{s}}-\nu _{\mathsf{s}}\mathsf{f}%
\alpha \\
\mu _{\mathsf{c}}[\alpha ]& =\mu _{\mathsf{c}}-\mathsf{f}\alpha \cdot \nu _{%
\mathsf{sc}}  \notag \\
\phi \lbrack \alpha ][j_{\mathsf{f}},j_{\mathsf{s}},j_{\mathsf{c}}]& =\exp [-%
\mathsf{f}(1+j_{\mathsf{f}})\alpha \cdot j_{\mathsf{s}}]\phi \lbrack j_{%
\mathsf{f}},j_{\mathsf{s}},j_{\mathsf{c}}]  \notag
\end{align}%
The performance of the hedge strategy and the fair price of the derivative
are determined using the first and second moments extracted from this
equation.

\begin{figure}[!t]
\centering
\includegraphics[width=0.95%
\linewidth]{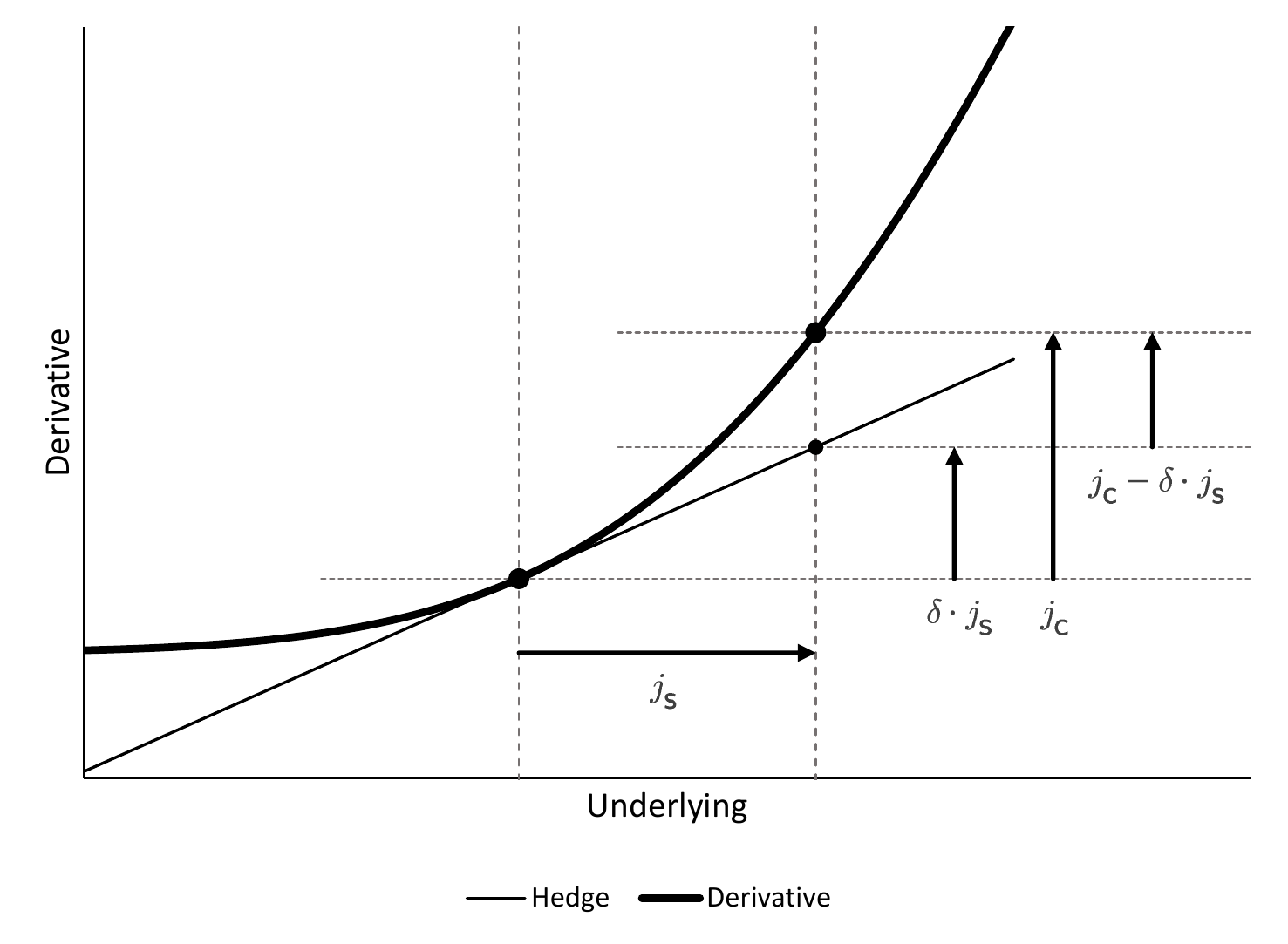}
\caption{The delta hedge matches the continuous price movements for the
derivative with an offsetting position in the underlying. Discontinuous
price movements invalidate the hedge when there is convexity in the
relationship between the underlying and derivative prices.}
\end{figure}

The covariances of the returns in the measure $\mathsf{\bar{z}}[\alpha ]$
are:%
\begin{equation}
\begin{bmatrix}
\mathsf{V}_{\mathsf{p}}[\alpha ] & \mathsf{C}_{\mathsf{pa}}[\alpha ] \\ 
\mathsf{C}_{\mathsf{pa}}^{t}[\alpha ] & \mathsf{V}_{\mathsf{a}}[\alpha ]%
\end{bmatrix}%
=\mathsf{f}^{2}%
\begin{bmatrix}
\mathsf{\dot{V}}_{\mathsf{s}}[\alpha ] & \mathsf{\dot{C}}_{\mathsf{sc}%
}[\alpha ] \\ 
\mathsf{\dot{C}}_{\mathsf{sc}}^{t}[\alpha ] & \mathsf{\dot{V}}_{\mathsf{c}%
}[\alpha ]%
\end{bmatrix}%
dt+O[dt^{2}]
\end{equation}%
where the variance rates are given by:%
\begin{align}
\mathsf{\dot{V}}_{\mathsf{s}}[\alpha ]& =\nu _{\mathsf{s}}+\int (1+j_{%
\mathsf{f}})^{2}j_{\mathsf{s}}^{2}\phi \lbrack \alpha ][j_{\mathsf{f}},j_{%
\mathsf{s}},j_{\mathsf{c}}]\,dj_{\mathsf{f}}\,dj_{\mathsf{s}}\,dj_{\mathsf{c}%
} \\
\mathsf{\dot{V}}_{\mathsf{c}}[\alpha ]& =\nu _{\mathsf{c}}+\int (1+j_{%
\mathsf{f}})^{2}j_{\mathsf{c}}^{2}\phi \lbrack \alpha ][j_{\mathsf{f}},j_{%
\mathsf{s}},j_{\mathsf{c}}]\,dj_{\mathsf{f}}\,dj_{\mathsf{s}}\,dj_{\mathsf{c}%
}  \notag
\end{align}%
and the covariance rate is given by:%
\begin{equation}
\mathsf{\dot{C}}_{\mathsf{sc}}[\alpha ]=\nu _{\mathsf{sc}}+\int (1+j_{%
\mathsf{f}})^{2}j_{\mathsf{s}}j_{\mathsf{c}}\phi \lbrack \alpha ][j_{\mathsf{%
f}},j_{\mathsf{s}},j_{\mathsf{c}}]\,dj_{\mathsf{f}}\,dj_{\mathsf{s}}\,dj_{%
\mathsf{c}}
\end{equation}%
The optimal hedge strategy minimises the variance for the return on the
hedged derivative. Quantifying performance using the equivalent measure with
parameters $\alpha $, the optimal hedge weights are:%
\begin{equation}
\beta =\mathsf{\dot{V}}_{\mathsf{s}}[\alpha ]^{-1}\mathsf{\dot{C}}_{\mathsf{%
sc}}[\alpha ]+O[dt]
\end{equation}%
This strategy does not necessarily eliminate the market risk of the hedged
derivative. The residual variance of the hedged portfolio is calculated as:%
\begin{equation}
\mathsf{v}=\mathsf{f}^{2}(\mathsf{\dot{V}}_{\mathsf{c}}[\alpha ]-\mathsf{%
\dot{V}}_{\mathsf{s}}[\alpha ]^{-1}\mathsf{\dot{C}}_{\mathsf{sc}}[\alpha
]\cdot \mathsf{\dot{C}}_{\mathsf{sc}}[\alpha ])\,dt+O[dt]
\end{equation}%
including contributions from the continuous and discontinuous components of
the price diffusion. The hedge strategy strikes a balance between these
contributions, optimising against both the infinitesimal price moves of the
continuous diffusion and the discrete price moves of the discontinuous
diffusion according to their relative likelihoods.

The continuous contribution dominates when $J^{2}\Phi $ is small, where $J$
is the maximum size of jumps in the support of the frequency density and $%
\Phi $ is the total frequency. The optimal hedge expands as:%
\begin{equation}
\beta =\delta +\gamma +O[dt]
\end{equation}%
The contribution $\delta $ from the continuous diffusion is the delta hedge
against infinitesimal changes in the prices, and this is corrected by the
adjustment $\gamma $ that accounts for the discontinuous diffusion:%
\begin{align}
\delta & =\nu _{\mathsf{s}}^{-1}\nu _{\mathsf{sc}} \\
\gamma & =\int (1+j_{\mathsf{f}})^{2}(j_{\mathsf{c}}-\delta \cdot j_{\mathsf{%
s}})(\nu _{\mathsf{s}}^{-1}j_{\mathsf{s}})\phi \lbrack \alpha ][j_{\mathsf{f}%
},j_{\mathsf{s}},j_{\mathsf{c}}]\,dj_{\mathsf{f}}\,dj_{\mathsf{s}}\,dj_{%
\mathsf{c}}+O[(J^{2}\Phi )^{2}]  \notag
\end{align}%
Price discontinuities corrupt the delta hedge strategy when there is
convexity in the relationship between underlying and derivative, requiring a
correction to the strategy in proportion to the deviation $(j_{\mathsf{c}%
}-\delta \cdot j_{\mathsf{s}})$ from the linear approximation. The delta
hedge does not depend on the parameters $\alpha $ that adjust the economic
measure, nor is it impacted by the volatility of the funding ratio.
Sensitivities to these elements are only introduced when there is
indeterminacy in the size and direction of the underlying and derivative
price jumps.

The means of the returns in the measure $\mathsf{\bar{z}}[\alpha ]$ are:%
\begin{equation}
\begin{bmatrix}
\mathsf{M}_{\mathsf{p}}[\alpha ] \\ 
\mathsf{M}_{\mathsf{a}}[\alpha ]%
\end{bmatrix}%
=\mathsf{f}%
\begin{bmatrix}
\mathsf{\dot{M}}_{\mathsf{s}}[\alpha ] \\ 
\mathsf{\dot{M}}_{\mathsf{c}}[\alpha ]%
\end{bmatrix}%
dt+O[dt^{2}]
\end{equation}%
where the mean rates are given by:%
\begin{align}
\mathsf{\dot{M}}_{\mathsf{s}}[\alpha ]& =\mu _{\mathsf{s}}[\alpha ]+\nu _{%
\mathsf{fs}}+\int (1+j_{\mathsf{f}})j_{\mathsf{s}}\phi \lbrack \alpha ][j_{%
\mathsf{f}},j_{\mathsf{s}},j_{\mathsf{c}}]\,dj_{\mathsf{f}}\,dj_{\mathsf{s}%
}\,dj_{\mathsf{c}} \\
\mathsf{\dot{M}}_{\mathsf{c}}[\alpha ]& =\mu _{\mathsf{c}}[\alpha ]+\nu _{%
\mathsf{fc}}+\int (1+j_{\mathsf{f}})j_{\mathsf{c}}\phi \lbrack \alpha ][j_{%
\mathsf{f}},j_{\mathsf{s}},j_{\mathsf{c}}]\,dj_{\mathsf{f}}\,dj_{\mathsf{s}%
}\,dj_{\mathsf{c}}  \notag
\end{align}%
The price model derived from the maximum entropy principle assumes the mean
of the derivative return is zero in the price measure that calibrates the
parameters $\alpha $ so that the means of the underlying returns are also
zero:%
\begin{align}
\mathsf{\dot{M}}_{\mathsf{s}}[\alpha ]& =0 \\
\mathsf{\dot{M}}_{\mathsf{c}}[\alpha ]& =0  \notag
\end{align}%
The calibration constraint is solved for the parameters that are then used
in the price equation for the derivative. These expressions simplify when
the underlying price diffusion is continuous, so that the frequency density
takes the form:%
\begin{equation}
\phi \lbrack j_{\mathsf{f}},j_{\mathsf{s}},j_{\mathsf{c}}]=\delta \lbrack j_{%
\mathsf{s}}]\phi \lbrack j_{\mathsf{f}},j_{\mathsf{c}}]
\end{equation}%
In this case, the solution for the parameters is entered into the condition
for the derivative return to generate the price equation:%
\begin{equation}
\mu _{\mathsf{c}}+\nu _{\mathsf{fc}}+\int (1+j_{\mathsf{f}})j_{\mathsf{c}%
}\phi \lbrack j_{\mathsf{f}},j_{\mathsf{c}}]\,dj_{\mathsf{f}}\,dj_{\mathsf{c}%
}=\nu _{\mathsf{s}}^{-1}\nu _{\mathsf{sc}}\cdot (\mu _{\mathsf{s}}+\nu _{%
\mathsf{fs}})
\end{equation}%
identifying the drift of the derivative price.

The assumption of continuous diffusion for the underlying prices can be
reasonable in benign market conditions. Liquidity is a desirable property of
market hedge instruments, and the price transparency and trading volumes on
exchanges facilitate continuous hedging. In stressed conditions, or for
hedging with less liquid instruments, the economic model should reflect the
potential difficulties with continuous hedging by incorporating a
discontinuous component for the underlying price diffusion. The calibration
parameters are then determined from the underlying drift condition via the
Newton-Raphson scheme.

\subsection{Stochastic volatility}

In the previous section, the drift conditions for the underlying and
derivative returns are derived from the maximum entropy principle on the
assumption of continuous margin settlement. The approach is demonstrated in
this section by considering in detail the economic model:%
\begin{align}
& \frac{\log \mathsf{\bar{z}}\bullet \exp [k_{\mathsf{s}}\,d\mathsf{s}+k_{%
\boldsymbol{\sigma }}\cdot d\boldsymbol{\sigma }]}{dt}= \\
& \hspace{1cm}k_{\mathsf{s}}\mu _{\mathsf{s}}+k_{\boldsymbol{\sigma }}\cdot
\mu _{\boldsymbol{\sigma }}+\frac{1}{2}k_{\mathsf{s}}^{2}\nu _{\mathsf{s}}+%
\frac{1}{2}k_{\boldsymbol{\sigma }}\cdot \nu _{\boldsymbol{\sigma }}k_{%
\boldsymbol{\sigma }}+k_{\mathsf{s}}k_{\boldsymbol{\sigma }}\cdot \nu _{%
\mathsf{s}\boldsymbol{\sigma }}  \notag \\
& \hspace{1cm}+\int (\exp [k_{\mathsf{s}}j_{\mathsf{s}}+k_{\boldsymbol{%
\sigma }}\cdot j_{\boldsymbol{\sigma }}]-1)\phi \lbrack j_{\mathsf{s}},j_{%
\boldsymbol{\sigma }}]\,dj_{\mathsf{s}}\,dj_{\boldsymbol{\sigma }}+O[dt] 
\notag
\end{align}%
describing the evolution in the economic measure of a single underlying
price $\mathsf{s}$ and an additional state vector $\boldsymbol{\sigma }$
used to model its volatility. Funding convexity is neglected by setting $%
\mathsf{f}=1$, and the mean rates $\mu _{\mathsf{s}}$ and $\mu _{\boldsymbol{%
\sigma }}$, covariance rates $\nu _{\mathsf{s}}$, $\nu _{\boldsymbol{\sigma }%
}$ and $\nu _{\mathsf{s}\boldsymbol{\sigma }}$, and jump frequency $\phi
\lbrack j_{\mathsf{s}},j_{\boldsymbol{\sigma }}]$ are all assumed to be
functions of time $t$ and the state variables $\mathsf{s}$ and $\boldsymbol{%
\sigma }$. Many popular models for pricing derivatives are included in this
setup.

\begin{description}
\item[Black-Scholes-Merton model:] This model assumes a stationary evolution
for the underlying price with constant volatility:%
\begin{align}
& \frac{\log \mathsf{\bar{z}}\bullet \exp [k_{\mathsf{s}}\,d\mathsf{s}/%
\mathsf{s}]}{dt}= \\
& \hspace{1cm}k_{\mathsf{s}}\mu +\frac{1}{2}k_{\mathsf{s}}^{2}\sigma
^{2}+\int_{j=-1}^{\infty }(\exp [k_{\mathsf{s}}j]-1)\psi \lbrack j]\,dj+O[dt]
\notag
\end{align}%
The model parameters are the mean rate $\mu $, the volatility rate $\sigma $%
, and the frequency of jumps $\psi \lbrack j]$ for the underlying price.

\item[Heston model:] This model assumes a stationary evolution for the
underlying price with stochastic volatility following a correlated
square-root process:%
\begin{align}
& \frac{\log \mathsf{\bar{z}}\bullet \exp [k_{\mathsf{s}}\,d\mathsf{s}/%
\mathsf{s}+k_{\boldsymbol{\nu }}\,d\boldsymbol{\nu }]}{dt}= \\
& \hspace{1cm}k_{\mathsf{s}}\mu +k_{\boldsymbol{\nu }}\kappa (\theta -%
\boldsymbol{\nu })+\frac{1}{2}k_{\mathsf{s}}^{2}\boldsymbol{\nu }+\frac{1}{2}%
k_{\boldsymbol{\nu }}^{2}\xi ^{2}\boldsymbol{\nu }+k_{\mathsf{s}}k_{%
\boldsymbol{\nu }}\rho \xi \boldsymbol{\nu }+O[dt]  \notag
\end{align}%
The additional state variable $\boldsymbol{\nu }$ is interpreted as the
instantaneous variance of the underlying price. The model parameters are the
mean rate $\mu $ of the underlying price, the mean reversion rate $\kappa $
and reversion level $\theta $ of the variance, the volatility of variance $%
\xi $, and the correlation $\rho $ between the price and its variance.

\item[SABR\ model:] This model assumes the underlying price has constant
elasticity of variance with stochastic volatility following a correlated
lognormal process:%
\begin{align}
& \frac{\log \mathsf{\bar{z}}\bullet \exp [k_{\mathsf{s}}\,d\mathsf{s}+k_{%
\boldsymbol{\sigma }}\,d\boldsymbol{\sigma }+k_{\boldsymbol{\alpha }}\,d%
\boldsymbol{\alpha }+k_{\boldsymbol{\beta }}\,d\boldsymbol{\beta }+k_{%
\boldsymbol{\rho }}\,d\boldsymbol{\rho }]}{dt}= \\
& \hspace{1cm}\frac{1}{2}k_{\mathsf{s}}^{2}\boldsymbol{\sigma }^{2}\mathsf{s}%
^{2\boldsymbol{\beta }}+\frac{1}{2}k_{\boldsymbol{\sigma }}^{2}\boldsymbol{%
\alpha }^{2}\boldsymbol{\sigma }^{2}+k_{\mathsf{s}}k_{\boldsymbol{\sigma }}%
\boldsymbol{\rho \alpha \sigma }^{2}\mathsf{s}^{\boldsymbol{\beta }}+O[dt] 
\notag
\end{align}%
The additional state variables are the instantaneous volatility $\boldsymbol{%
\sigma }$ of the underlying price, the volatility of volatility $\boldsymbol{%
\alpha }$, the elasticity of variance $\boldsymbol{\beta }$, and the
correlation $\boldsymbol{\rho }$ between the price and its volatility.
\end{description}

The maximum entropy principle locates the price measure $\mathsf{\hat{z}}$
among the equivalent measures $\mathsf{\bar{z}}[\alpha ]$ parametrised by
the weight $\alpha $ for the underlying return. The evolution of the state
variables in this measure is modelled by:%
\begin{align}
& \frac{\log \mathsf{\bar{z}}[\alpha ]\bullet \exp [k_{\mathsf{s}}\,d\mathsf{%
s}+k_{\boldsymbol{\sigma }}\cdot d\boldsymbol{\sigma }]}{dt}= \\
& \hspace{1cm}k_{\mathsf{s}}\mu _{\mathsf{s}}[\alpha ]+k_{\boldsymbol{\sigma 
}}\cdot \mu _{\boldsymbol{\sigma }}[\alpha ]+\frac{1}{2}k_{\mathsf{s}%
}^{2}\nu _{\mathsf{s}}+\frac{1}{2}k_{\boldsymbol{\sigma }}\cdot \nu _{%
\boldsymbol{\sigma }}k_{\boldsymbol{\sigma }}+k_{\mathsf{s}}k_{\boldsymbol{%
\sigma }}\cdot \nu _{\mathsf{s}\boldsymbol{\sigma }}  \notag \\
& \hspace{1cm}+\int (\exp [k_{\mathsf{s}}j_{\mathsf{s}}+k_{\boldsymbol{%
\sigma }}\cdot j_{\boldsymbol{\sigma }}]-1)\phi \lbrack \alpha ][j_{\mathsf{s%
}},j_{\boldsymbol{\sigma }}]\,dj_{\mathsf{s}}\,dj_{\boldsymbol{\sigma }%
}+O[dt]  \notag
\end{align}%
with adjusted parameters:%
\begin{align}
\mu _{\mathsf{s}}[\alpha ]& =\mu _{\mathsf{s}}-\alpha \nu _{\mathsf{s}} \\
\mu _{\boldsymbol{\sigma }}[\alpha ]& =\mu _{\boldsymbol{\sigma }}-\alpha
\nu _{\mathsf{s}\boldsymbol{\sigma }}  \notag \\
\phi \lbrack \alpha ][j_{\mathsf{s}},j_{\boldsymbol{\sigma }}]& =\exp
[-\alpha j_{\mathsf{s}}]\phi \lbrack j_{\mathsf{s}},j_{\boldsymbol{\sigma }}]
\notag
\end{align}%
These adjustments modify the mean rates and jump frequency of the price
diffusion via the control variable $\alpha $, which is used to calibrate the
price measure to the underlying returns.

Market completeness for the derivative depends on the availability of hedge
securities and the nature of their joint price dynamics. In the following,
hedge strategies are considered for the derivative with price $\mathsf{c}%
=c[t,\mathsf{s},\boldsymbol{\sigma }]$ expressed as a function of time and
the state variables. Hedging with only the underlying neglects the
contributions to volatility from the additional state variables, leaving
residual risk in the portfolio. Hedge performance is improved if the market
includes options on the underlying that mark the state variables.

In the incomplete hedging scenario, the returns on the hedge and derivative
securities are:%
\begin{equation}
\mathsf{R}_{\mathsf{p}}=d\mathsf{s\hspace{1cm}R}_{\mathsf{a}}=d\mathsf{c}
\end{equation}%
The means of the returns in the equivalent measure $\mathsf{\bar{z}}[\alpha
] $ are:%
\begin{equation}
\begin{bmatrix}
\mathsf{M}_{\mathsf{p}}[\alpha ] \\ 
\mathsf{M}_{\mathsf{a}}[\alpha ]%
\end{bmatrix}%
=%
\begin{bmatrix}
\mathsf{\dot{M}}_{\mathsf{s}}[\alpha ] \\ 
\mathsf{\dot{M}}_{\mathsf{c}}[\alpha ]%
\end{bmatrix}%
dt+O[dt^{2}]
\end{equation}%
where the mean rates are given by:%
\begin{align}
\mathsf{\dot{M}}_{\mathsf{s}}[\alpha ]=\,& \mu _{\mathsf{s}}[\alpha ]+\int
j_{\mathsf{s}}\phi \lbrack \alpha ][j_{\mathsf{s}},j_{\boldsymbol{\sigma }%
}]\,dj_{\mathsf{s}}\,dj_{\boldsymbol{\sigma }} \\
\mathsf{\dot{M}}_{\mathsf{c}}[\alpha ]=\,& \frac{\partial c}{\partial t}+%
\frac{\partial c}{\partial \mathsf{s}}\mu _{\mathsf{s}}[\alpha ]+\frac{%
\partial c}{\partial \boldsymbol{\sigma }}\cdot \mu _{\boldsymbol{\sigma }%
}[\alpha ]+\frac{1}{2}\frac{\partial ^{2}c}{\partial \mathsf{s}^{2}}\nu _{%
\mathsf{s}}+\frac{1}{2}\func{tr}[\frac{\partial ^{2}c}{\partial \boldsymbol{%
\sigma }^{2}}\nu _{\boldsymbol{\sigma }}]+\frac{\partial ^{2}c}{\partial 
\mathsf{s}\partial \boldsymbol{\sigma }}\cdot \nu _{\mathsf{s}\boldsymbol{%
\sigma }}  \notag \\
& +\int j_{\mathsf{c}}\phi \lbrack \alpha ][j_{\mathsf{s}},j_{\boldsymbol{%
\sigma }}]\,dj_{\mathsf{s}}\,dj_{\boldsymbol{\sigma }}  \notag
\end{align}%
The jump $j_{\mathsf{c}}$ for the derivative price, appearing in the
integrand of the discontinuous contribution, is defined in terms of the
jumps $j_{\mathsf{s}}$ and $j_{\boldsymbol{\sigma }}$ for the underlying
price and state variables:%
\begin{equation}
j_{\mathsf{c}}=c[t,\mathsf{s}+j_{\mathsf{s}},\boldsymbol{\sigma }+j_{%
\boldsymbol{\sigma }}]-c[t,\mathsf{s},\boldsymbol{\sigma }]
\end{equation}

The equation for the fair price of the derivative is obtained by setting the
means of the underlying and derivative returns to zero in the price measure $%
\mathsf{\hat{z}}=\mathsf{\bar{z}}[\alpha ]$, with the first condition used
to calibrate the weight $\alpha $ that is then applied in the second
condition to determine the derivative price:%
\begin{align}
0=\,& \mu _{\mathsf{s}}-\alpha \nu _{\mathsf{s}}+\int j_{\mathsf{s}}\exp
[-\alpha j_{\mathsf{s}}]\phi \lbrack j_{\mathsf{s}},j_{\boldsymbol{\sigma }%
}]\,dj_{\mathsf{s}}\,dj_{\boldsymbol{\sigma }} \\
0=\,& \frac{\partial c}{\partial t}+\frac{\partial c}{\partial \boldsymbol{%
\sigma }}\cdot (\mu _{\boldsymbol{\sigma }}-\alpha \nu _{\mathsf{s}%
\boldsymbol{\sigma }})+\frac{1}{2}\frac{\partial ^{2}c}{\partial \mathsf{s}%
^{2}}\nu _{\mathsf{s}}+\frac{1}{2}\func{tr}[\frac{\partial ^{2}c}{\partial 
\boldsymbol{\sigma }^{2}}\nu _{\boldsymbol{\sigma }}]+\frac{\partial ^{2}c}{%
\partial \mathsf{s}\partial \boldsymbol{\sigma }}\cdot \nu _{\mathsf{s}%
\boldsymbol{\sigma }}  \notag \\
& +\int (j_{\mathsf{c}}-\frac{\partial c}{\partial \mathsf{s}}j_{\mathsf{s}%
})\exp [-\alpha j_{\mathsf{s}}]\phi \lbrack j_{\mathsf{s}},j_{\boldsymbol{%
\sigma }}]\,dj_{\mathsf{s}}\,dj_{\boldsymbol{\sigma }}  \notag
\end{align}%
These expressions simplify when the underlying price diffusion is
continuous, so that the frequency density takes the form:%
\begin{equation}
\phi \lbrack j_{\mathsf{s}},j_{\boldsymbol{\sigma }}]=\delta \lbrack j_{%
\mathsf{s}}]\phi \lbrack j_{\boldsymbol{\sigma }}]
\end{equation}%
In this case, the calibration constraint is solved by $\alpha =\mu _{\mathsf{%
s}}/\nu _{\mathsf{s}}$, and the price equation becomes:%
\begin{align}
0=\,& \frac{\partial c}{\partial t}+\frac{\partial c}{\partial \boldsymbol{%
\sigma }}\cdot (\mu _{\boldsymbol{\sigma }}-\frac{\mu _{\mathsf{s}}}{\nu _{%
\mathsf{s}}}\nu _{\mathsf{s}\boldsymbol{\sigma }})+\frac{1}{2}\frac{\partial
^{2}c}{\partial \mathsf{s}^{2}}\nu _{\mathsf{s}}+\frac{1}{2}\func{tr}[\frac{%
\partial ^{2}c}{\partial \boldsymbol{\sigma }^{2}}\nu _{\boldsymbol{\sigma }%
}]+\frac{\partial ^{2}c}{\partial \mathsf{s}\partial \boldsymbol{\sigma }}%
\cdot \nu _{\mathsf{s}\boldsymbol{\sigma }} \\
& +\int j_{\mathsf{c}}\phi \lbrack j_{\boldsymbol{\sigma }}]\,dj_{%
\boldsymbol{\sigma }}  \notag
\end{align}%
The partial differential-integral equation for the fair price of the
derivative is solved against the boundary conditions provided by the
terminal settlements, as specified in the derivative contract.

Now suppose that the hedge market includes options on the underlying, with
prices $\mathsf{o}=o[t,\mathsf{s},\boldsymbol{\sigma }]$ expressed as
functions of time and the state variables. Further assume that the options
mark the additional state variables, with implied volatility function $%
\boldsymbol{\sigma }=\sigma \lbrack t,\mathsf{s},\mathsf{o}]$ satisfying:%
\begin{align}
\sigma \lbrack t,\mathsf{s},o[t,\mathsf{s},\boldsymbol{\sigma }]]& =%
\boldsymbol{\sigma } \\
o[t,\mathsf{s},\sigma \lbrack t,\mathsf{s},\mathsf{o}]]& =\mathsf{o}  \notag
\end{align}%
These options enable the hedging of all the state variables, thereby
reducing the residual risk of the hedged derivative.

In the complete hedging scenario, the returns on the hedge and derivative
securities are:%
\begin{equation}
\mathsf{R}_{\mathsf{p}}=%
\begin{bmatrix}
d\mathsf{s} \\ 
d\mathsf{o}%
\end{bmatrix}%
\mathsf{\hspace{1cm}R}_{\mathsf{a}}=d\mathsf{c}
\end{equation}%
The covariances of the returns in the economic measure $\mathsf{\bar{z}}$
are:%
\begin{equation}
\begin{bmatrix}
\mathsf{V}_{\mathsf{p}} & \mathsf{C}_{\mathsf{pa}} \\ 
\mathsf{C}_{\mathsf{pa}}^{t} & \mathsf{V}_{\mathsf{a}}%
\end{bmatrix}%
=%
\begin{bmatrix}
\mathsf{\dot{V}}_{\mathsf{s}} & \mathsf{\dot{C}}_{\mathsf{so}}^{t} & \mathsf{%
\dot{C}}_{\mathsf{sc}} \\ 
\mathsf{\dot{C}}_{\mathsf{so}} & \mathsf{\dot{V}}_{\mathsf{o}} & \mathsf{%
\dot{C}}_{\mathsf{oc}} \\ 
\mathsf{\dot{C}}_{\mathsf{sc}} & \mathsf{\dot{C}}_{\mathsf{oc}}^{t} & 
\mathsf{\dot{V}}_{\mathsf{c}}%
\end{bmatrix}%
dt+O[dt^{2}]
\end{equation}%
where the variance rates are given by:%
\begin{align}
\mathsf{\dot{V}}_{\mathsf{s}}& =\nu _{\mathsf{s}}+\int j_{\mathsf{s}%
}^{2}\phi \lbrack j_{\mathsf{s}},j_{\boldsymbol{\sigma }}]\,dj_{\mathsf{s}%
}\,dj_{\boldsymbol{\sigma }} \\
\mathsf{\dot{V}}_{\mathsf{o}}& =(\frac{\partial o}{\partial \mathsf{s}}%
)^{2}\nu _{\mathsf{s}}+\frac{\partial o}{\partial \boldsymbol{\sigma }}\cdot
\nu _{\boldsymbol{\sigma }}\frac{\partial o}{\partial \boldsymbol{\sigma }}+2%
\frac{\partial o}{\partial \mathsf{s}}\frac{\partial o}{\partial \boldsymbol{%
\sigma }}\cdot \nu _{\mathsf{s}\boldsymbol{\sigma }}+\int j_{\mathsf{o}%
}^{2}\phi \lbrack j_{\mathsf{s}},j_{\boldsymbol{\sigma }}]\,dj_{\mathsf{s}%
}\,dj_{\boldsymbol{\sigma }}  \notag \\
\mathsf{\dot{V}}_{\mathsf{c}}& =(\frac{\partial c}{\partial \mathsf{s}}%
)^{2}\nu _{\mathsf{s}}+\frac{\partial c}{\partial \boldsymbol{\sigma }}\cdot
\nu _{\boldsymbol{\sigma }}\frac{\partial c}{\partial \boldsymbol{\sigma }}+2%
\frac{\partial c}{\partial \mathsf{s}}\frac{\partial c}{\partial \boldsymbol{%
\sigma }}\cdot \nu _{\mathsf{s}\boldsymbol{\sigma }}+\int j_{\mathsf{c}%
}^{2}\phi \lbrack j_{\mathsf{s}},j_{\boldsymbol{\sigma }}]\,dj_{\mathsf{s}%
}\,dj_{\boldsymbol{\sigma }}  \notag
\end{align}%
and the covariance rates are given by:%
\begin{align}
\mathsf{\dot{C}}_{\mathsf{so}}& =\frac{\partial o}{\partial \mathsf{s}}\nu _{%
\mathsf{s}}+\frac{\partial o}{\partial \boldsymbol{\sigma }}\cdot \nu _{%
\mathsf{s}\boldsymbol{\sigma }}+\int j_{\mathsf{s}}j_{\mathsf{o}}\phi
\lbrack j_{\mathsf{s}},j_{\boldsymbol{\sigma }}]\,dj_{\mathsf{s}}\,dj_{%
\boldsymbol{\sigma }} \\
\mathsf{\dot{C}}_{\mathsf{sc}}& =\frac{\partial c}{\partial \mathsf{s}}\nu _{%
\mathsf{s}}+\frac{\partial c}{\partial \boldsymbol{\sigma }}\cdot \nu _{%
\mathsf{s}\boldsymbol{\sigma }}+\int j_{\mathsf{s}}j_{\mathsf{c}}\phi
\lbrack j_{\mathsf{s}},j_{\boldsymbol{\sigma }}]\,dj_{\mathsf{s}}\,dj_{%
\boldsymbol{\sigma }}  \notag \\
\mathsf{\dot{C}}_{\mathsf{oc}}& =\frac{\partial o}{\partial \mathsf{s}}\frac{%
\partial c}{\partial \mathsf{s}}\nu _{\mathsf{s}}+\frac{\partial o}{\partial 
\boldsymbol{\sigma }}\cdot \nu _{\boldsymbol{\sigma }}\frac{\partial c}{%
\partial \boldsymbol{\sigma }}+(\frac{\partial o}{\partial \mathsf{s}}\frac{%
\partial c}{\partial \boldsymbol{\sigma }}+\frac{\partial o}{\partial 
\boldsymbol{\sigma }}\frac{\partial c}{\partial \mathsf{s}})\cdot \nu _{%
\mathsf{s}\boldsymbol{\sigma }}+\int j_{\mathsf{o}}j_{\mathsf{c}}\phi
\lbrack j_{\mathsf{s}},j_{\boldsymbol{\sigma }}]\,dj_{\mathsf{s}}\,dj_{%
\boldsymbol{\sigma }}  \notag
\end{align}%
The jumps $j_{\mathsf{o}}$ and $j_{\mathsf{c}}$ for the option and
derivative prices, appearing in the integrand of the discontinuous
contribution, are defined in terms of the jumps $j_{\mathsf{s}}$ and $j_{%
\boldsymbol{\sigma }}$ for the underlying price and state variables:%
\begin{align}
j_{\mathsf{o}}& =o[t,\mathsf{s}+j_{\mathsf{s}},\boldsymbol{\sigma }+j_{%
\boldsymbol{\sigma }}]-o[t,\mathsf{s},\boldsymbol{\sigma }] \\
j_{\mathsf{c}}& =c[t,\mathsf{s}+j_{\mathsf{s}},\boldsymbol{\sigma }+j_{%
\boldsymbol{\sigma }}]-c[t,\mathsf{s},\boldsymbol{\sigma }]  \notag
\end{align}

Hedging with the underlying portfolio $\beta _{\mathsf{s}}$ and the option
portfolio $\beta _{\mathsf{o}}$, the residual variance for the hedged
derivative is:%
\begin{align}
\mathsf{v}=\,& ((\frac{\partial c}{\partial \mathsf{s}}-\beta _{\mathsf{s}%
}-\beta _{\mathsf{o}}\cdot \frac{\partial o}{\partial \mathsf{s}})^{2}\nu _{%
\mathsf{s}} \\
& +(\frac{\partial c}{\partial \boldsymbol{\sigma }}-\frac{\partial o}{%
\partial \boldsymbol{\sigma }}\beta _{\mathsf{o}})\cdot \nu _{\boldsymbol{%
\sigma }}(\frac{\partial c}{\partial \boldsymbol{\sigma }}-\frac{\partial o}{%
\partial \boldsymbol{\sigma }}\beta _{\mathsf{o}})  \notag \\
& +2(\frac{\partial c}{\partial \mathsf{s}}-\beta _{\mathsf{s}}-\beta _{%
\mathsf{o}}\cdot \frac{\partial o}{\partial \mathsf{s}})(\frac{\partial c}{%
\partial \boldsymbol{\sigma }}-\frac{\partial o}{\partial \boldsymbol{\sigma 
}}\beta _{\mathsf{o}})\cdot \nu _{\mathsf{s}\boldsymbol{\sigma }}  \notag \\
& +\int (j_{\mathsf{c}}-\beta _{\mathsf{s}}j_{\mathsf{s}}-\beta _{\mathsf{o}%
}\cdot j_{\mathsf{o}})^{2}\phi \lbrack j_{\mathsf{s}},j_{\boldsymbol{\sigma }%
}]\,dj_{\mathsf{s}}\,dj_{\boldsymbol{\sigma }})\,dt+O[dt^{2}]  \notag
\end{align}%
The choice of hedge portfolio then depends on the objectives of the investor
and their level of access to the hedge market. Using only the underlying,
the delta hedge is:%
\begin{equation}
\beta _{\mathsf{s}}=\frac{\partial c}{\partial \mathsf{s}}\mathsf{\hspace{1cm%
}}\beta _{\mathsf{o}}=0
\end{equation}%
This strategy offsets the impact on the derivative price of continuous moves
in the underlying price, but neglects the impact from the volatility of
state variables. The residual variance achieved by the strategy is:%
\begin{equation}
\mathsf{v}=(\frac{\partial c}{\partial \boldsymbol{\sigma }}\cdot \nu _{%
\boldsymbol{\sigma }}\frac{\partial c}{\partial \boldsymbol{\sigma }}+\int
(j_{\mathsf{c}}-\beta _{\mathsf{s}}j_{\mathsf{s}})^{2}\phi \lbrack j_{%
\mathsf{s}},j_{\boldsymbol{\sigma }}]\,dj_{\mathsf{s}}\,dj_{\boldsymbol{%
\sigma }})\,dt+O[dt^{2}]
\end{equation}%
Performance of the delta hedge is improved by including options to offset
the continuous moves in the state variables:%
\begin{equation}
\beta _{\mathsf{s}}=\frac{\partial c}{\partial \mathsf{s}}-(\frac{\partial o%
}{\partial \boldsymbol{\sigma }})^{-1}\frac{\partial o}{\partial \mathsf{s}}%
\cdot \frac{\partial c}{\partial \boldsymbol{\sigma }}\mathsf{\hspace{1cm}}%
\beta _{\mathsf{o}}=(\frac{\partial o}{\partial \boldsymbol{\sigma }})^{-1}%
\frac{\partial c}{\partial \boldsymbol{\sigma }}
\end{equation}%
This strategy removes the contribution from the continuous covariance of the
underlying and option prices, leaving only the discontinuous term:%
\begin{equation}
\mathsf{v}=(\int (j_{\mathsf{c}}-\beta _{\mathsf{s}}j_{\mathsf{s}}-\beta _{%
\mathsf{o}}\cdot j_{\mathsf{o}})^{2}\phi \lbrack j_{\mathsf{s}},j_{%
\boldsymbol{\sigma }}]\,dj_{\mathsf{s}}\,dj_{\boldsymbol{\sigma }%
})\,dt+O[dt^{2}]
\end{equation}%
The prize-winning observation is that the continuous contribution to the
residual variance can be completely eliminated when there are sufficient
hedge securities available to match against the volatile state variables.
Moreover, the discontinuous contribution is small if the price jumps broadly
align with the linear approximation, so that the residual $(j_{\mathsf{c}%
}-\beta _{\mathsf{s}}j_{\mathsf{s}}-\beta _{\mathsf{o}}\cdot j_{\mathsf{o}})$
is negligible. Large market disruptions are commonly associated with the
breakdown of normal relationships between prices, however, and the hedged
derivative is exposed to the risk of discontinuities if these moves deviate
from the tangent relationship. Variance reduction could be extended to
account for such regime switches, adapting the hedge strategy to balance the
impact from continuous and discontinuous market moves, but no strategy can
be completely successful at eliminating the market risks if the direction of
price moves is indeterminate.

The assumptions that validate the delta hedging strategy, namely that margin
is settled continuously and the underlying price moves are themselves
continuous, are at best an approximation to real hedging activity. The price
model accommodates the impact of discrete market moves within the timescale
of trading by admitting discontinuity in the underlying diffusion,
substituting risk elimination with portfolio optimisation as the unifying
principle for pricing.

It should be noted that none of the core economic principles that are the
basis for the price model hold true even in the most elementary markets. The
principle of replicability disregards consideration of trading volume that
may impact price via illiquidity or economy of scale. The principle of
no-arbitrage denies the existence of arbitrage, which is certainly possible
over short horizons and has been observed to persist even on the scale of
days or months in exceptional circumstances. Arguably the most pernicious of
the three, the principle of economic equivalence has the potential to
corrupt the price model through invalid economic suppositions, inviting
technical assumptions whose mathematical attractiveness can mask the
realities of market dynamics.

\section{Literature review}

Understanding the relationship between the economic measure and the price
measure has been the fundamental question of mathematical finance ever since
Bachelier first applied stochastic calculus in his pioneering thesis \cite%
{Bachelier1900}. In this thesis, Bachelier develops a model for the
logarithm of the security price as Brownian motion around its equilibrium,
deriving expressions for options that would not look unfamiliar today.
Methods of portfolio optimisation originated in the work of Markowitz \cite%
{Markowitz1952} and Sharpe \cite{Sharpe1964} using Gaussian statistics for
the returns, generalised by later authors to allow more sophisticated
distributional assumptions and measures of utility. These approaches are
embedded in the economic measure, explaining the origin of price as the
equilibrium of market activity uncovering the expectations of participants.

The impact of dynamic hedging on price was first recognised in the articles
by Black and Scholes \cite{Black1973} and Merton \cite{Merton1973}. Adopting
the framework devised by Bachelier, these authors observe that the option
return is replicated exactly by a strategy that continuously offsets the
delta of the option price to the underlying price. The theory matured with
the work of Harrison and Pliska \cite{Harrison1981,Harrison1983}, providing
a precise statement of the conditions for market completeness and the
martingale property of price in a measure equivalent to the economic
measure. The discipline has since expanded in numerous directions, with
significant advances for term structure and default modelling and numerical
methods for complex derivative structures.

While market evidence for discounting basis existed earlier, the basis
widening that occurred as a result of the Global Financial Crisis of
2007-2008 motivated research into funding and its impact on discounting.
Work by Johannes and Sundaresan \cite{Johannes2007}, Fujii and Takahashi 
\cite{Fujii2011}, Piterbarg and Antonov \cite%
{Piterbarg2010,Piterbarg2012,Antonov2014}, Henrard \cite{Henrard2014},
McCloud \cite{McCloud2013a} and others established the theoretical
justification and practical application of collateral discounting.

Entropy methods have found application across the range of mathematical
finance, including portfolio optimisation \cite{Philippatos1972} and
derivative pricing \cite{Buchen1996,Gulko1999} -- see also the review essay 
\cite{Zhou2013} which includes further references. The article \cite%
{Frittelli2000} by Frittelli proposes the minimal relative entropy measure
as a solution to the problem of pricing in incomplete markets, and links
this solution with the maximisation of expected exponential utility. As a
measure of disorder, entropy performs a role similar to variance but is
better suited to the real distributions of market returns. Proponents of the
use of entropy justify the approach by appeal to the modelling of
information flows in dynamical systems and the analogy with thermodynamics.

Most approaches to derivative pricing begin with the assumption of
continuous settlement, and stochastic calculus is an essential ingredient
for these developments. The representation of the stochastic differential
equation used in this article follows the discoveries of L\'{e}vy \cite%
{Levy1934}, Khintchine \cite{Khintchine1937} and It\^{o} \cite{Ito1941}. The
technical requirements of the L\'{e}vy-Khintchine representation can be
found in these articles and other standard texts in probability theory. The
change from economic measure to price measure implied by the maximum entropy
principle extends the result from Girsanov \cite{Girsanov1960} to include
the scaling adjustment of the jump density in addition to the drift
adjustment.

The example models referenced in the section on stochastic volatility are
the Black-Scholes-Merton model \cite{Black1973,Merton1976}, the Heston model 
\cite{Heston1993} and the SABR model \cite{Hagan2002}, all commonly used
models in quantitative finance.

\end{document}